\documentclass[aps,onecolumn,preprint,groupedaddress,superscriptaddress,showpacs]{revtex4-1}

\usepackage{graphicx}
\usepackage{color}
\usepackage{amsmath}
\usepackage{amsfonts}
\usepackage{amssymb}
\usepackage{dcolumn}
\usepackage{hyperref}
\usepackage{bbold}
\usepackage{cancel}
\begin{document}
   \title{Impact of embedding on predictability of failure-recovery dynamics in networks}

 \author{L. B\"{o}ttcher}
  \email{lucasb@ethz.ch}
\author{M. Lukovi\'c}
\email{lukovicm@ethz.ch}
\author{J. Nagler}
\email{jnagler@ethz.ch}
 \affiliation{ETH Zurich, Wolfgang-Pauli-Strasse 27, CH-8093 Zurich,
Switzerland}
\author{S. Havlin}
\affiliation{Center for Polymer Studies and Department of Physics, Boston University, Boston, Massachusetts 02215, USA}
\affiliation{Department of Physics, Bar-Ilan University, 52900 Ramat-Gan, Israel}
 \author{H. J. Herrmann}
  \affiliation{ETH Zurich, Wolfgang-Pauli-Strasse 27, CH-8093 Zurich,
Switzerland}  
 \affiliation{
Departamento de F\'isica, Universidade
Federal do Cear\'a, 60451-970 Fortaleza, Cear\'a, Brazil}
\date{\today}
\begin{abstract}
Failure, damage spread and recovery crucially underlie many spatially embedded networked systems 
ranging from transportation structures  to the human body. 
Here we study the interplay between spontaneous damage, induced failure and recovery in both
embedded and non-embedded networks.
In our model the network's components follow three realistic processes that capture these features: 
(i) spontaneous failure of a component independent of the neighborhood (internal failure),
(ii) failure induced by failed neighboring nodes (external failure) and 
(iii) spontaneous recovery of a component. 
We identify a metastable domain in the global network phase diagram spanned by the model's control parameters 
where dramatic hysteresis effects and random switching between two coexisting states are observed.
The loss of predictability due to these effects depend on the characteristic link length of the embedded system. 
For the Euclidean lattice in particular, hysteresis and switching only occur in an extremely narrow region of the parameter space compared to random networks.
We develop a unifying theory %for embedded and random networks 
which links the dynamics of our model to contact processes.  
Our unifying framework may help 
to better understand predictability and controllability in spatially embedded and random networks 
where spontaneous recovery of components can mitigate
spontaneous failure and damage spread in the global network. 
\end{abstract}
\maketitle
\section{Introduction}
\label{sec:intro}
Failure, damage spread and recovery crucially underlie many spatially embedded networked systems ranging from transportation structures to the human body \cite{verma16,barabasi11,barthelemy11}.
 Advances in the study of networks have led to important progress in understanding resilience and controllability in terms of the interaction between topology and various underlying spreading dynamics \cite{reuven2000,reka2000,centola05,buldyrev10,Helbing13,bashan13}. In the case of a simple contagion or contact process such as an epidemic, it is possible for the disease to spread from a single infected source to other neighboring individuals. On the other hand, many phenomena such as the diffusion of innovations \cite{coleman57,rogers2010diffusion}, political mobilization \cite{chwe99}, viral marketing \cite{leskovec07} and coordination games \cite{easley2010} are characterized by a complex contagion where nodes need to be connected to multiple sources in order to induce a change of their state \cite{granovetter78,macy07}.
%The case where a single contact between nodes is sufficient to allow spreading is referred to as simple contagion or contact process.
%However, 
In addition to this induced transition, individuals may %also tend to 
spontaneously change their opinion or banks can spontaneously fail \cite{may08, haldane11}.
%and we therefore explicitly 
%Our model includes this spontaneous node failure which
%This calls to investigate the influence of an additional noise term 
%accounting for such behavior. 
%This 
%leads to %more 
%a rich dynamics characterized by a metastable domain in the parameter space where switching between two coexisting 
%global 
%states is observed.

%The proper theoretical description and classification enables us to develop a better intuition for the dynamics. 
%We use this model to study 
The consequences of the interplay between spontaneous damage, induced failure and recovery %for resilience and controllability
of components in spatially embedded systems are crucial for systemic risk \cite{lorenz09}, predictability and controllability but have not yet been systematically explored.
Many real-world networks such as power grids, computer networks and social networks are embedded in Euclidean space \cite{barthelemy11}.
% and our analysis
 %will
 %Impact of spatial regularity on predictability of failure-recovery networks
We here show how the process of
embedding and the related characteristic link length impact the predictability of failure-recovery dynamics in networks.

  %shows that this feature has great impact on the system's 
 % robustness. % in the following manner: 
Our model is based on three fundamental processes (i) spontaneous failure independent of the neighborhood (internal failure), (ii) failure induced by failed neighboring nodes if their number exceeds a threshold (external failure) and (iii) spontaneous recovery (see Fig. \ref{fig:model}). The interplay between these three processes results in a phase diagram with a metastable regime where hysteresis and switching between two coexisting states are observed  \cite{majdandzic14}. 
  %The failure-recovery 
  %dynamics based on the fundamental processes exhibits
  %metastability 
  %  where \lb{hysteresis and} switching between two coexisting states \lb{are} observed \cite{majdandzic14}. %The sudden and, in terms of controllability, not desirable
     %This, %\lb{in terms of controllability}, undesired 
     %phase switching implies} %to 
     In technological systems, hysteresis effects % imply discontinuous transitions 
     %which
      might be
     potentially harmful since slight changes of the system's control parameters can entail drastic and abrupt transitions
      from a seemingly globally stable state  
       to macroscopic inactivity or large-scale outage \cite{achlioptas09,araujo10,nagler11,nagler12,schroeder13,cho13,chen13,chen132,Helbing13,
       chen14,boettcher14,souza15,boettcher16,saberi16,schroeder16}. 
       Hysteresis and spontaneous switching between coexisting states in multistable dynamical systems has received great attention for processes ranging from decision making \cite{turalska09}, multistable perception \cite{ditzinger89, moreno07, Atwal14} over fluid phase transitions \cite{kesselring13}, protein folding and unfolding \cite{ding2002} to chemical oscillations \cite{kim2002}, % and
       magnetic systems \cite{binder84} and
       human sleep stages \cite{bashan12}.
%
 %or %a potentially completely
       %global damaged \cite{Helbing13}. 
       We therefore propose here that the extent of the
metastable regime in the parameter space of the phase diagram can be regarded as a measure
of predictability. The larger the metastable regime the lower is the predictability of the
system.
Based on this measure we find that the networks' predictability strongly increases with its regularity. 
%embedded systems are substantially more predictable in the presence of failure and recovery dynamics -- compared to non-embedded random networks. 
In particular, for the Euclidean lattice,  hysteresis only occurs in a very small range of the spontaneous failure rate
-- compared to random networks with the same average number of neighbors.

Our analytical approach is based on mapping
 %and we also show that 
 the dynamics
 % should be seen as a 
 to a generalized contact process where a certain minimum number of failed neighboring nodes is necessary to activate the induced failure \cite{tome15}. 
 This strongly suggests
 %As a result, we argue 
 that the dynamics does not belong to the Ising universality class as 
 %pointed out 
 conjectured earlier \cite{majdandzic14}. %Furthermore, we describe that our model is not only capturing the more general contact process dynamics but also 
 In addition, we show that our model system is inherently linked to 
%we demonstrate that our model system describes 
 complex contagion phenomena \cite{granovetter78,macy07} and %with %where together with so-called %the 
 %displays 
 %a phase space equivalent to 
 cusp catastrophes \cite{ludwig78, zeeman79, strogatz14}. Our unifying framework
        %Our framework, however, unifies
        for random %networks 
and %spatially 
partially
embedded networks helps to better understand predictability and controllability in % spatially embedded systems 
systems where spontaneous recovery can mitigate %or even counterbalance 
spontaneous failure and damage spread.

\section{Materials and Methods}
\label{sec:model}

\begin{figure}
\begin{minipage}{0.49\textwidth}
\centering
\includegraphics[width=\textwidth]{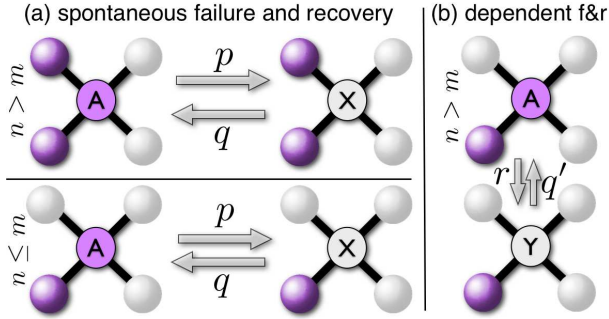}
\end{minipage}
  \caption{\textbf{Model.} (a) Spontaneous failure ($A\rightarrow X$) and spontaneous recovery  ($X\rightarrow A$) takes place with rates $p$ and $q$, respectively. 
  (b) A node may also fail (become inactive) dependent on its neighborhood,
   if too few {\em active} nodes $n\le m$ sustain the node's activity ($A\rightarrow Y$ with rate $r$).
  In addition, a failed node $Y$ recovers ($Y\rightarrow A$) with rate $q'$. 
  Illustration for $m=1$. Active nodes are purple.} 
 \label{fig:model}
\end{figure}
%
%
%
%To investigate and better understand the interplay between spontaneous damage, induced failure and recovery, 
We study a modified version of the failure-recovery model proposed in Ref. \cite{majdandzic14}. 
Specifically, 
we consider a fully rate based kinetic Monte Carlo model \cite{gillespie76, gillespie77} instead of, as previously, 
assuming \emph{fixed recovery times $\tau\neq 0$ and $\tau'=1$}.
The system's components (i.e.\ nodes) are regarded as either active (not damaged) or inactive (failed).
 %and our
 The  dynamics is based on three fundamental processes: 
 (i) a node spontaneously fails in a time interval $dt$ with probability 
 $p \text{d}t$ (internal failure), 
 (ii) if fewer than or equal to $m$ nearest neighbors of a certain node are active, this node fails due to external causes with probability $r \text{d}t$ (external failure) and (iii) spontaneous recovery with probability $q \text{d}t$ (internal recovery) or probability $q' \text{d}t$ (external recovery). 
%We consider a fully rate based kinetic Monte Carlo model \cite{gillespie76, gillespie77} instead of, as previously, assuming \emph{fixed recovery times $\tau\neq 0$ and $\tau'=1$}.
 The threshold $m$, similar to threshold rules in complex contagion models \cite{granovetter78,watts02,lopez08} %decides 
 determines 
 if the neighborhood is 
 critically 
 damaged or healthy (Fig.\ \ref{fig:model}). 
  A low value of $m$ describes the case where %means that %describe spread where 
  a large number of infected neighbors is required in order to sustain the spread of an innovation, opinion or damage.
Hence, unlike in an epidemic, where a single infected neighbor % node
can infect 
 a susceptible node, %($m=k-1$), %gets infected if at least one neighboring node is infected,
 in
 %\jn{
% For
 complex contagion processes 
 %This differs from simple contagion in that, unlike a disease, it may not be possible for 
 %the innovation to
  %the 
  spread %(of an innovation, opinion or damage)
   %is only likely when
   requires  more than one infected neighbor. %s are infected. % ($k-m > 1$) . %(case $m<
  %after only one incident of contact with an infected neighbor.
 %On a regular random graph with degree $k$, for example, this condition is met if
 %at least 
% $m=k-1$.
 %More complex contagion processes such as failures of interconnected banks may be described by %exhibit 
 %lower thresholds, $m < k-1$.

Let $a(t)\in [0,1]$ denote the total fraction of %totally
 {\em failed} nodes and $z(t)=1-a(t)$ the fraction of \emph{active} ones.
 Thus, $a(t)=u_{\text{int}}(t)+u_{\text{ext}}(t)$ with $u_{\text{int}}(t)$ and $u_{\text{ext}}(t)$ being the fractions of internal and external failure respectively. The total fraction of failed nodes in the stationary state is referred to as $a_{st}$. %(\cdot)$. 
 For the derivation of the mean-field rate equations we assume perfect mixing and first concentrate on the internal failure dynamics. The rate equation of internally failed nodes is given by:
\begin{equation}
\frac{d u_{\text{int}}(t)}{d t}= p \left(1-a(t)\right)- q u_{\text{int}}(t),
\label{eq:internal_rate}
\end{equation}
where the first term accounts for the fact that active nodes internally fail with rate $p$ and the second term corresponds to the recovery of internally failed nodes.

A node is said to be located in a critically damaged neighborhood (CDN) if its number of active neighbors is smaller than or equal to $m$. External failure is only acting on nodes in a CDN. The probability that a node of degree $k$ is located in a CDN is $E_k=\sum_{j=0}^m  \binom {k} {k-j} a^{k-j} (1-a)^j$ \cite{majdandzic14}. Consequently, the time evolution of the external failure is described by:
\begin{equation}
\frac{d u_{\text{ext}}(t)}{d t}= r \sum_k f_k E_k \left(1-a(t)\right)- q' u_{\text{ext}}(t),
\label{eq:external_rate}
\end{equation}
with $f_k$ being the degree distribution. The first term describes the failure of active nodes in a CDN with rate $r$ and the second term accounts for recovery of externally failed nodes with rate $q'$.

%XXX
%TAKEN OUT DOES NOT RELATE TO THE INTRO OF THE MODEL BUT TO THE RESULTS / DISCUSSION
% For certain values in the parameter space we encounter closed orbits as shown in Appendix \ref{sec:oscillations}. We further describe the dynamics and connections to other models, in particular, Schl\"ogl's first (contact process) and second model and the relation to cusp catastrophes, in Appendix \ref{sec:connection}.
%
%
%

\section{Results}
\subsection{Time evolution and phase-switching in embedded systems}
\label{sec:switching}
\begin{figure}
\begin{minipage}{0.49\textwidth}
\centering
\includegraphics[width=\textwidth]{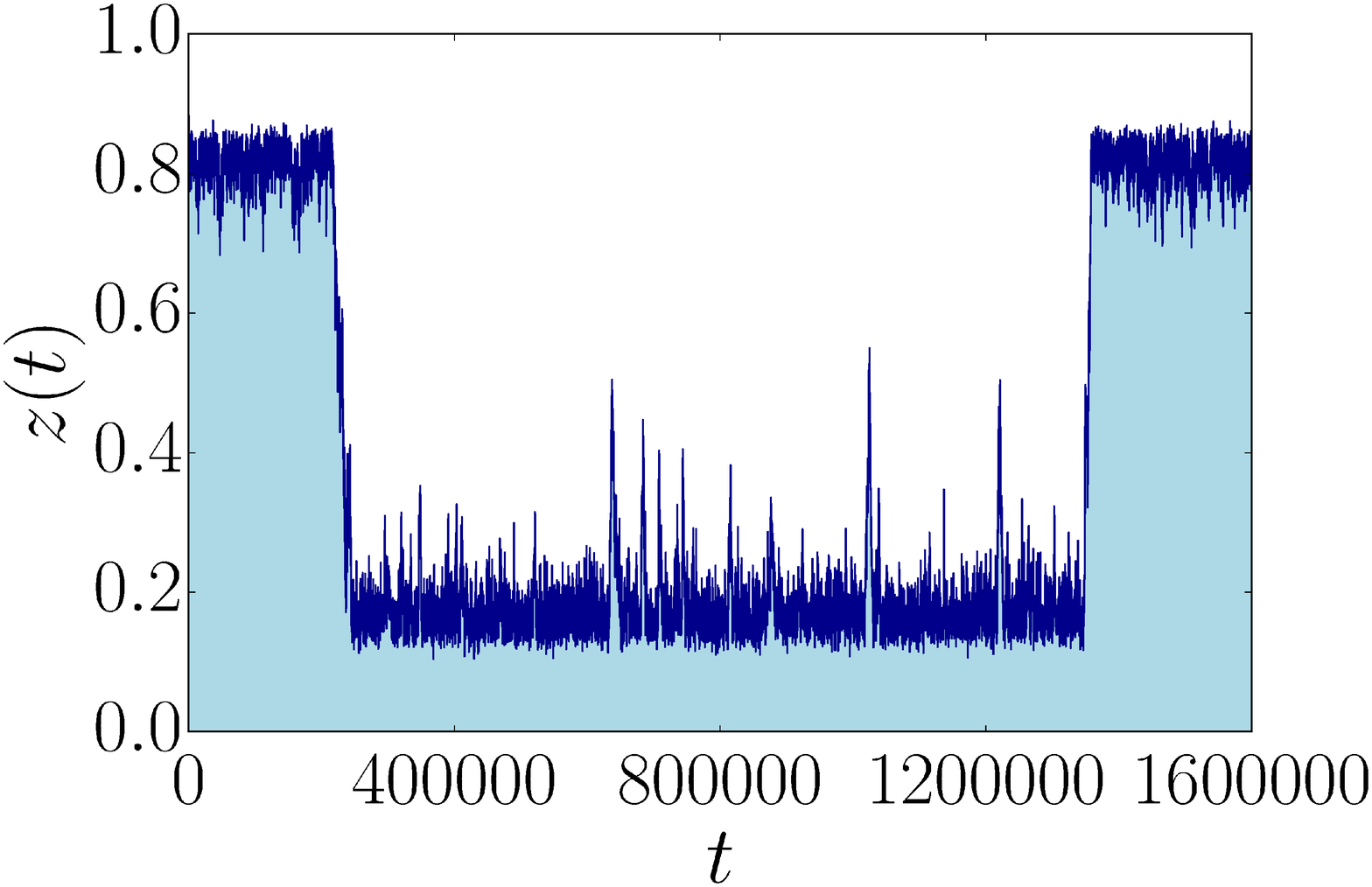}
\end{minipage}
\begin{minipage}{0.49\textwidth}
\centering
\includegraphics[width=\textwidth]{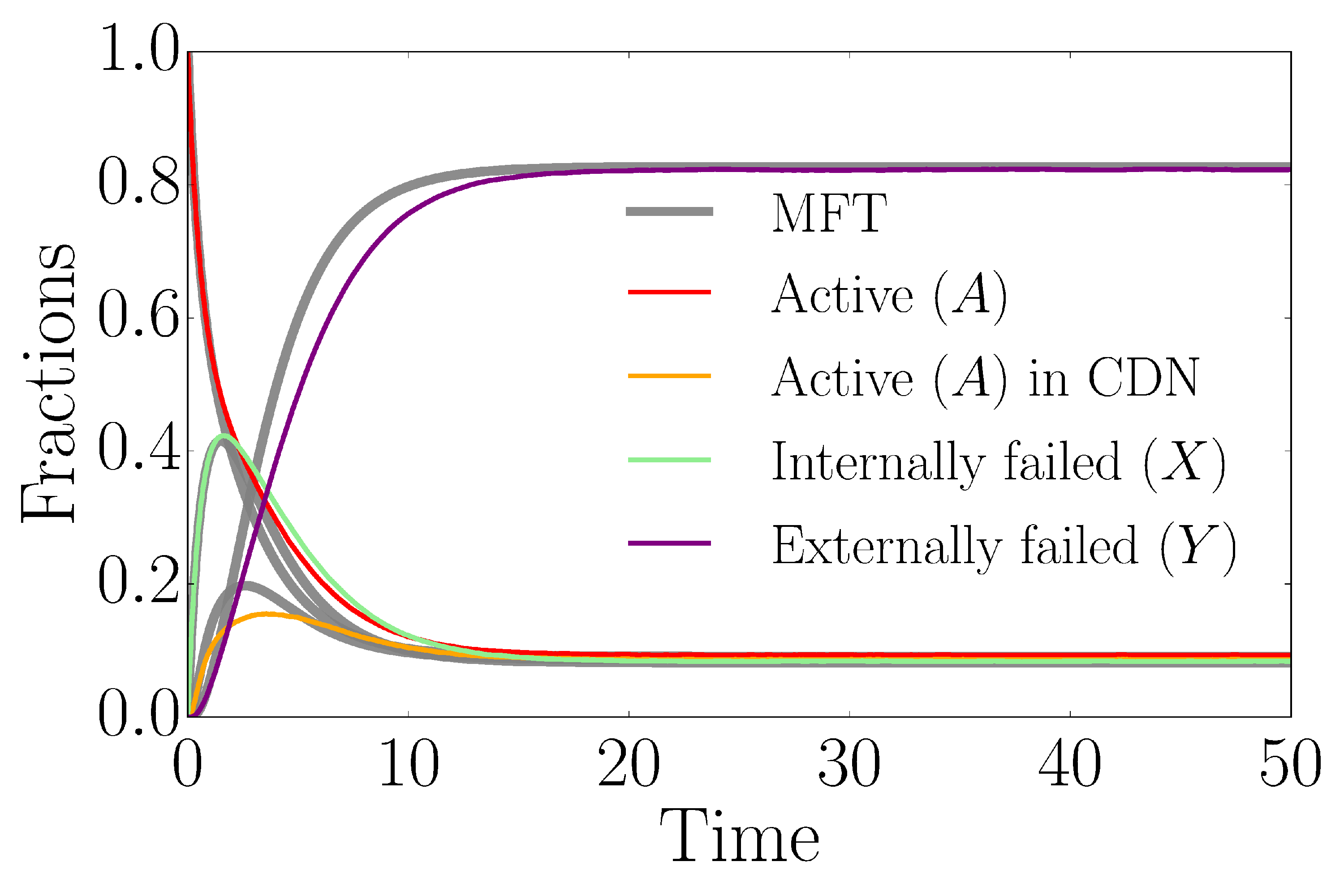}
\end{minipage}
  \caption{\textbf{Model dynamics on a square lattice.} 
  (left) Phase switching for $p=0.1065$, $r=0.95$, $q=1.0$, $q'=0.1$ and $m=1$ on a square lattice with $N=50\times 50$ nodes ($z$ the fraction of active nodes). (right) Time evolution of different model compartments, i.e. nodes in a certain
state (see Fig. \ref{fig:model}), with $p=0.9$, $r=0.95$, $q=1.0$, $q'=0.1$ and $m=1$ for a square lattice with $N=1024\times 1024$ nodes. All nodes are initially active, $z(0)=1$.} 
 \label{fig:dynamics}
\end{figure}
%
%
%
%With the %introduced
The coupled 
 mean-field rate equations Eqs. \eqref{eq:internal_rate} and \eqref{eq:external_rate} 
 determine the 
 %we 
 %are now able to 
 %relate the 
 time evolution of the dynamics as shown in  % to a theoretical description. 
 %We show the time evolution of different model compartments in a square lattice with $N=1024\times 1024$ nodes in
  Fig.\ \ref{fig:dynamics}. 
  %The mean-field theory describes the typical features of the underlying dynamics. 
  %In this example, all nodes are active in the beginning. 
  Internal failures first dominate the dynamics but after some time externally induced failures start becoming prevalent in the system.
   Interestingly, due to the system wide spread of the total failure after a transient phase, indicated by the small fraction of active nodes, 
  the relative abundance of the nodes susceptible to external (active in a CDN) and to internal failure (active)
  saturate at the same level.
  However, the process with the higher spreading rate (here external failure with $r/q'>p/q$) 
  soon
  %then 
  dominates the dynamics. 
  %That
  %\jn{CHECK:
  This explains the relatively small contribution of internal failure in this 
  parameter
   range
  which could not be observed employing %the %following the 
  %It also illustrates the approximate character of 
  the %previously introduced 
  mean-field theory %valid %for %the special case of $q=q'$
   of Ref.\ \cite{majdandzic14} which assumes that internal and external failure are %independent processes, 
    effectively decoupled processes (case $1\approx \tau'\ll \tau$). Therefore, our dynamical theory allows to analytically describe the
time evolution of the model's compartments, i.e. nodes in a certain
state.
  %WHAT ABOUT THE OTHER PAPER?
  %}
  %since both processes are stochastically dependent \cite{majdandzic14}.
%
%
%
\begin{figure}
\begin{minipage}{0.49\textwidth}
\centering
\includegraphics[width=\textwidth]{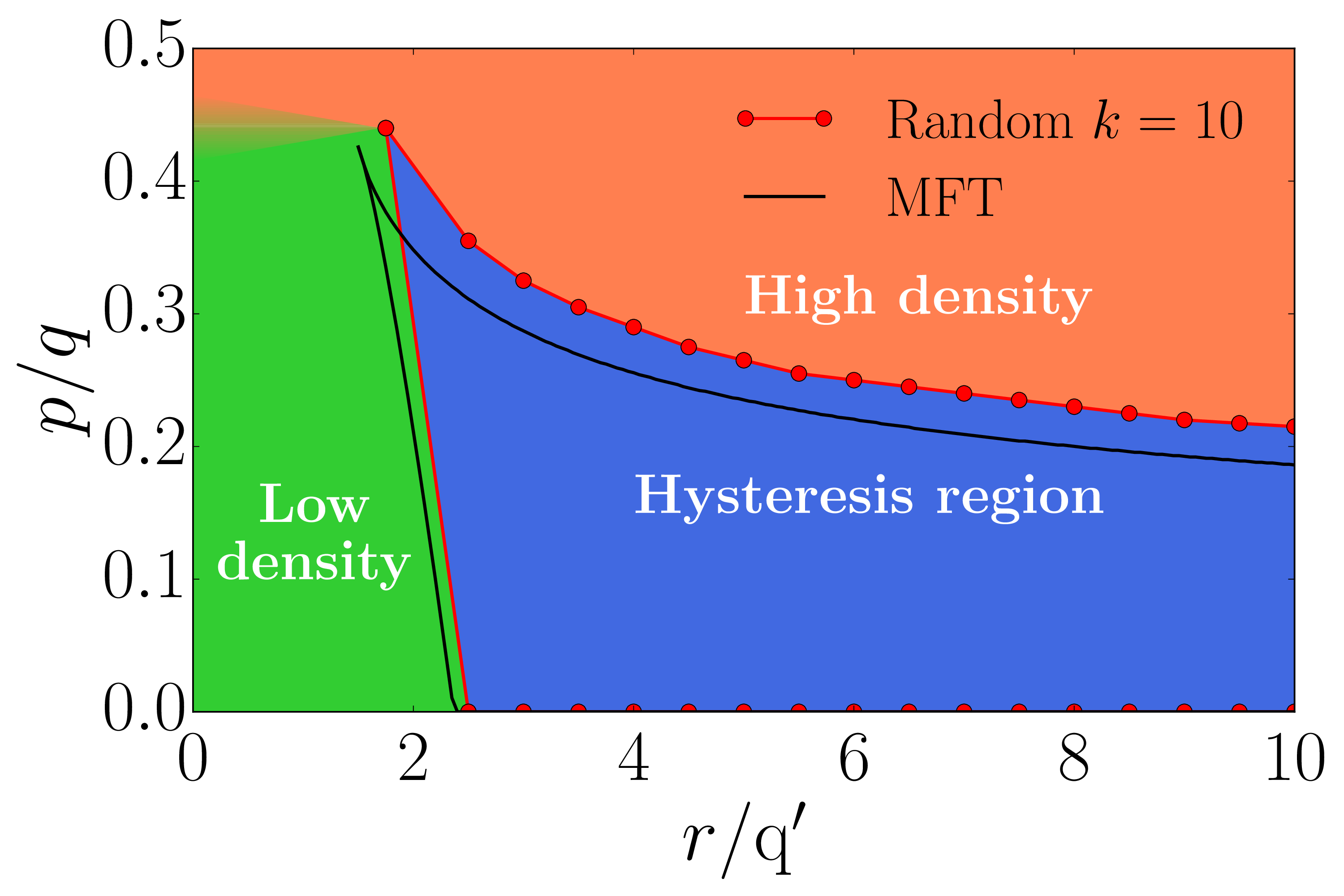}
\end{minipage}
\begin{minipage}{0.49\textwidth}
\centering
\includegraphics[width=\textwidth]{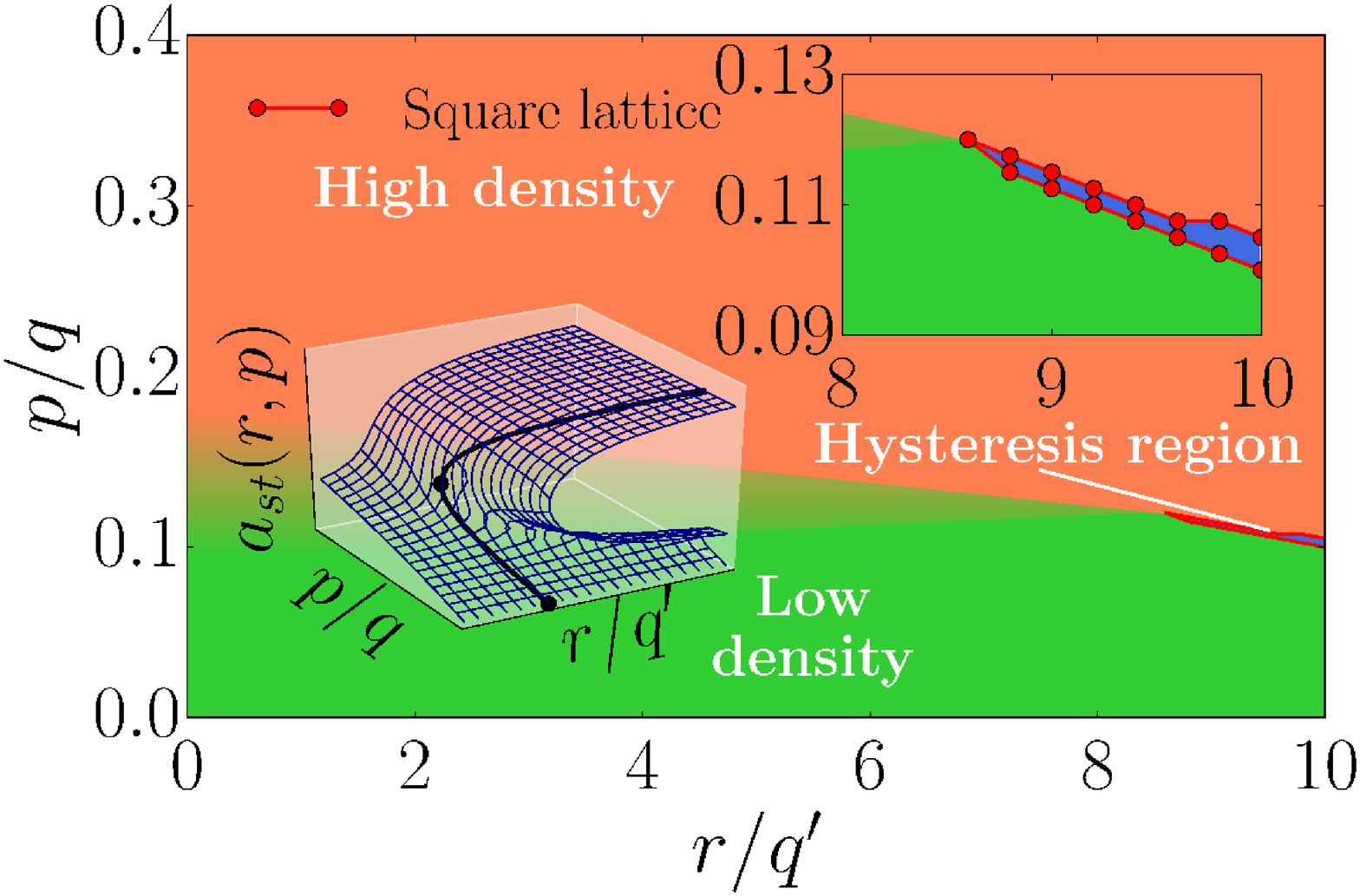}
\end{minipage}
  \caption{\textbf{Phase diagrams of a regular random graph with $k=10$ and a square lattice.} 
(left) The phase space of a regular random graph with $N=100,000$ nodes and $k=10$, $m=4$ (red dots) compared to the mean-field prediction (black lines).  
(right) The phase diagram of the square lattice (red spinodals) for $m=1$ obtained through simulations on a system with $N=2048\times 2048$ nodes.
Inset shows a blow-up of the
%In the inset in the right upper corner, we show a magnified 
hysteresis region. Compared to the regular random network (left), this region is very narrow. As an inset in the lower left corner, we show a typical cusp catastrophe surface \cite{zeeman79} (mean-field $k=4$, $m=1$) whose bifurcation lines (black) enclose the hysteresis region. For both plots the recovery rates are kept constant ($q=1.0$, $q'=0.1$) and $r$, $p$ are varied. %compared to the mean-field situation (black spinodals). The mean-field bifurcation point is given by $(r_0/q',p_0/q)=(3125/1296,19/81)$ and the one of the lattice is located at $(0.86(1),0.117(3))$. The simulations have been performed on a square lattice with $N=2048\times 2048$ nodes. For both plots the recovery rates are kept constant ($q=1.0$, $q'=0.1$) and $r$, $p$ are varied.
 } 
 \label{fig:phase_spaces_sl_rand}
\end{figure}
%
%
%

%The study of phase diagrams of random networks revealed a hysteresis region where phase-switching between two steady states has been observed \cite{majdandzic14}. %already said 
%We also found the 

For the Euclidean (square) lattice we observe 
phase-switching as shown in %phenomenon for the Euclidean lattice as a typical example for an embedded system. 
%In 
Fig. \ref{fig:dynamics} (left).
% (left) 
%we show a typical switching event in a lattice with $N=50\times 50$ nodes. 
The fraction of active nodes $z(t)$ undergoes rapid transitions between a phase of high and low activity. 
%Interestingly, 
%For the Euclidean lattice in particular, 
Hysteresis only occurs for large node-to-node spreading rates $r$ 
and a very narrow range of spontaneous failures rates $p$. 
Thus, we find the hysteresis region of
 the  lattice to be 
  much narrower than in 
   random graphs as shown in Fig. \ref{fig:phase_spaces_sl_rand}. 
   Inside this metastable domain rapid and unpredictable phase-switching occurs. In addition, crossing this region results in abrupt and dramatic transitions and it might not be possible to go back to the previous state following the same path. As an example, one can consider a nearly healthy population with a varying spontaneous infection rate which can cause the population to undergo a catastrophic transition to a highly infected state by crossing into the hysteresis region. Going back to the healthy state might not be as easy as just retracing the path followed before.
  In that sense, the extent of the hysteresis region in parameter space can be regarded as a measure for the
  predictability of the network's dynamics. The smaller the hysteresis region the less likely it is for the system to end up in this unpredictable situation. 
  However, the unpredictability in the hysteresis region is manifested in two ways: (1) In finite systems, random phase switching between two unstable states is observed where the mean of the random switching times
increases exponentially with system size \cite{majdandzic14}. 
(2) In the thermodynamic limit 
no switching is observed but the initial configuration and 
small random events in the initial temporal evolution of the system determine to which of the two stationary states the system converges. The latter behavior is characteristic for non-self-averaging spin glasses \cite{sherrington75,kirkpatrick78,sornette06}.
  As an inset in Fig. \ref{fig:phase_spaces_sl_rand}, we show the relation to cusp catastrophe surfaces accompanying the model's dynamics. Cusp catastrophes are a prominent example in catastrophe theory describing hysteresis and possible sudden transitions as a consequence of slightly varying control parameters with applications in population dynamics, mechanical and biological systems \cite{ludwig78, zeeman79, strogatz14}. The bifurcation lines enclose the hysteresis region and merge at the cusp point.
  %where a third-order polynomial with a vanishing second order term in the Taylor expansion characterize 
  %But in a cusp geometry the bifurcation curve loops back on itself, giving a second branch where this alternate solution itself loses stability, and will make a jump back to the original solution set. 
  The cusp point is a degenerate critical point 
  %Catastrophe theory describes degenerate critical points 
  where not just the first derivative, but also higher derivatives of the potential function vanish.
  %These are called the germs of the catastrophe geometries. 
  The degeneracy of this critical point can be unfolded by expanding the potential function as a Taylor series in small perturbations of the parameters $r,~p,$ and $a$ with a characteristic fourth-order polynomial \cite{zeeman79}.
   For a %more 
   detailed analytical treatment, see % in 
   Appendices \ref{sec:connection} and \ref{sec:critical_behavior}.
  We define the hysteresis areas enclosed by the bifurcation lines in this parameter range as $A_H^{MFT}$ (mean-field) and $A_H^{SL}$ (square lattice). %This area can be interpreted as a robustness measure and 
  The ratio $A_H^{MFT}:A_H^{SL}\approx 200:1 $ shows %. This result shows 
  %that large external spreading rates $r$ are necessary in the square lattice due to the dominance of local interactions. This means that the Euclidean lattice is substantially more robust against abrupt spontaneous and cascading \lb{failures compared to} random networks. 
that the Euclidean lattice is substantially more predictable in the presence of failure and recovery compared to random networks,
where non-local connections induce a faster damage spread.
For the square lattice, in contrast to random networks, 
failure cascades are only sustained for a large damage spreading rate r within a narrow region of the ratio $p/q$.  
  To obtain the phase diagram we studied the hysteresis behavior and fluctuations for different fixed values of $r$ by varying $p$. 
  More details about the Euclidean lattice and its critical behavior for different values of $m$ are described in Appendix \ref{sec:critical_behavior}. The model's dynamics is very rich and we show in \ref{sec:oscillations} that for certain values in the parameter space we encounter closed orbits. We further describe the dynamics and connections to other models, in particular, Schl\"ogl's first (contact process) and second models and the relation to cusp catastrophes, in Appendix \ref{sec:connection}.
\subsection{Spreading dynamics}
\label{sec:spreading}
\begin{figure}
\begin{minipage}{0.49\textwidth}
\centering
\includegraphics[width=0.55\textwidth]{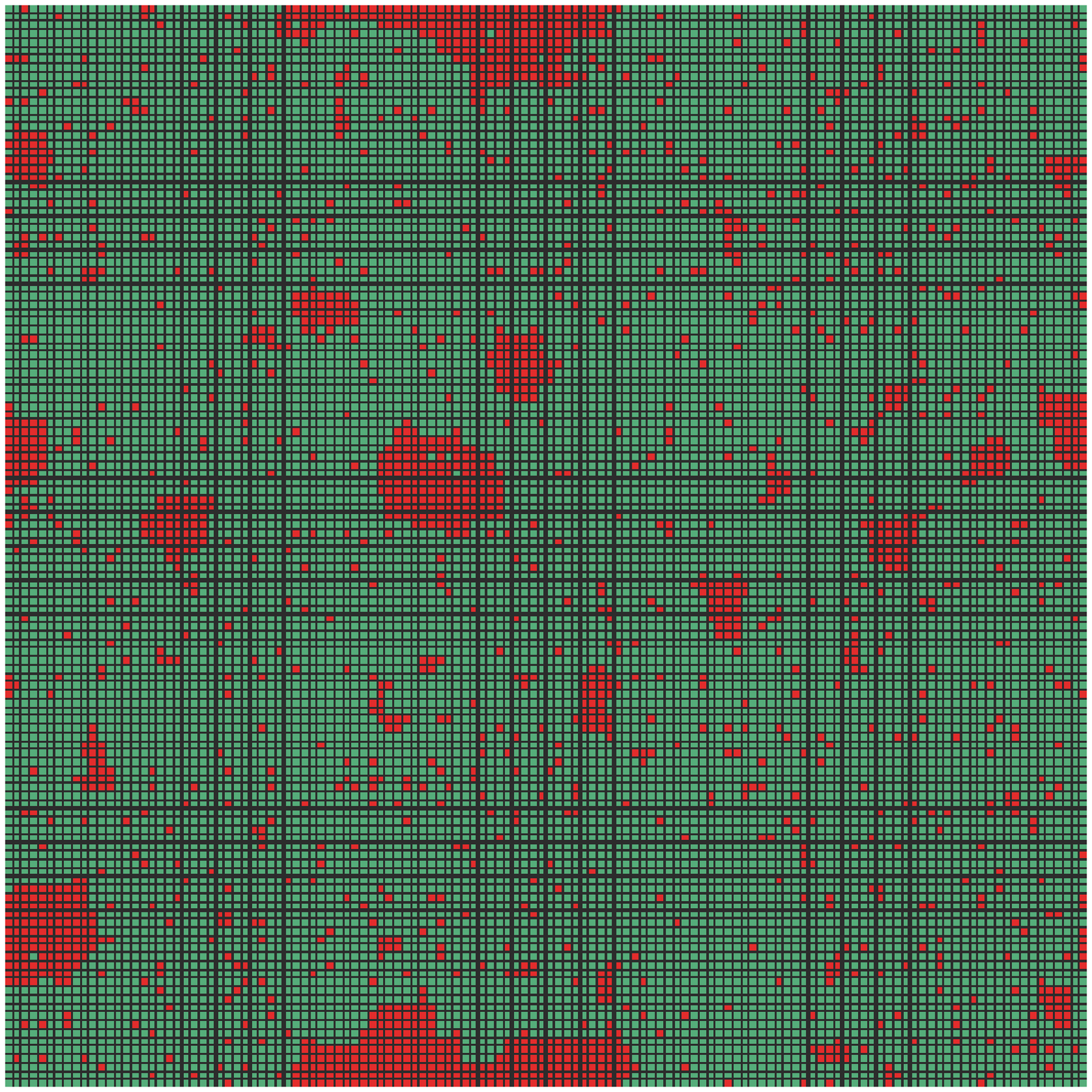}
\end{minipage}
\begin{minipage}{0.49\textwidth}
\centering
\includegraphics[width=0.55\textwidth]{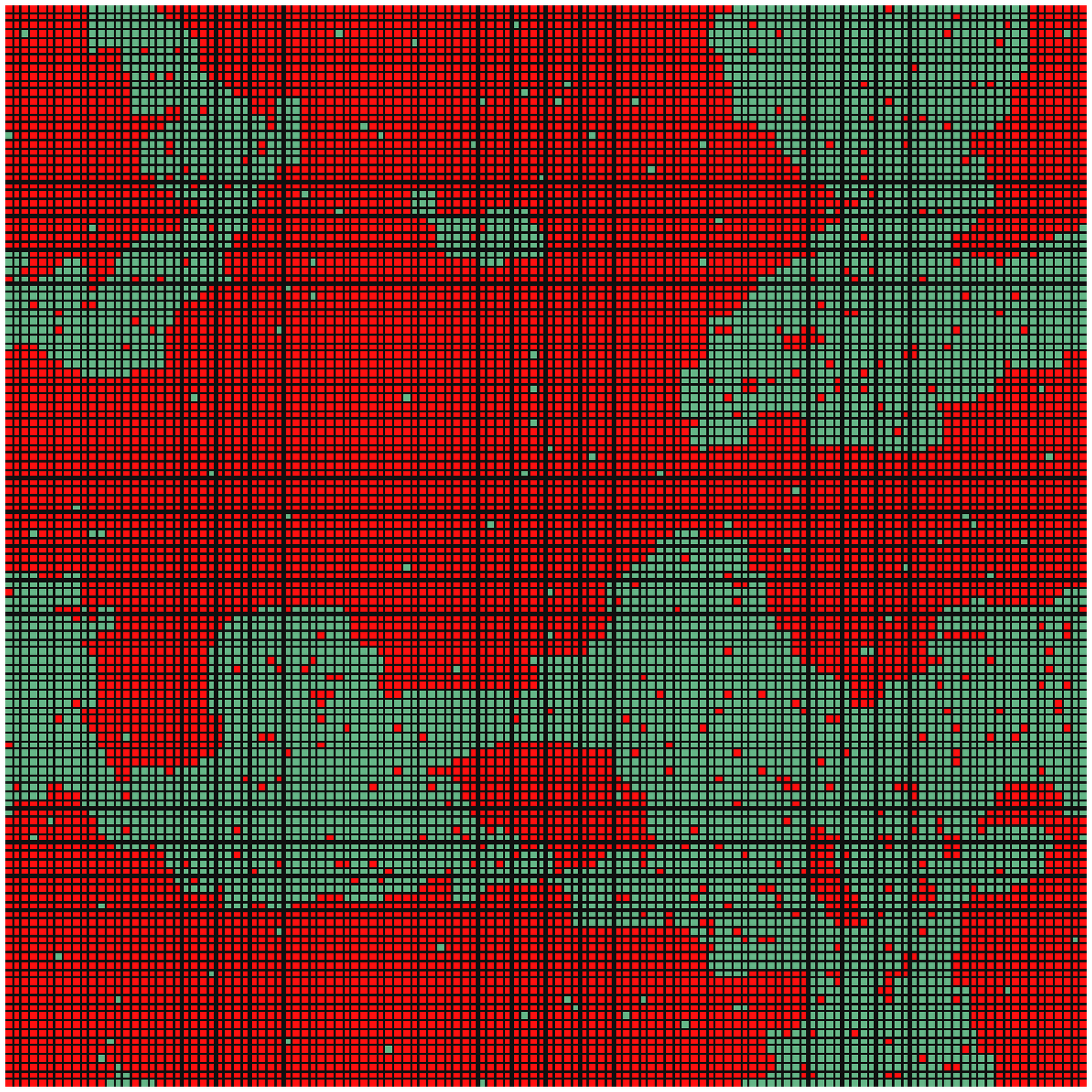}
\end{minipage}
  \caption{\textbf{Nucleation for vanishing spontaneous infection.} Simulation of the spontaneous recovery model with $p=0.05$, $q=1.0$, $r=10.0$, $q'=0.1$ and $m=1$ on a square lattice with $N=128\times 128$ nodes. (left) Initially, multiple spreading seeds of failed lattice sites (red) form due to spontaneous failure. (right) Contact dynamics (external failure) dominate and active sites (green) are displaced by failed ones (red). For further details we refer to the video version of the dynamics: \href{https://vimeo.com/163988456}{video 3 (vanishing spontaneous infection)}.} 
 \label{fig:spreading}
\end{figure}
%
%
%
%Different points in the phase space correspond to different spreading dynamics in the lattice. % would be bad if not
%In principal, %typo
%We distinguish situations where 
The dynamics can be %is 
driven by the field-like spontaneous failure term or %by %where 
the spread of failure can be triggered by the neighboring failed nodes. We first discuss the dynamics in the hysteresis region of a square lattice with $N=50 \times 50$ nodes, cf. Fig. \ref{fig:dynamics}. The two 
mechanisms
%transitions 
are illustrated in \href{https://vimeo.com/162425603}{video 1 (transition down)} \cite{video1} and \href{https://vimeo.com/162425733}{video 2 (transition up)} \cite{video2}. The spontaneous infection term, analogous to an external field, enables the dynamics to form multiple seeds from where the transitions might start. The regions invaded by different seeds expand and move on the lattice. Some of them merge and form larger clusters of active or failed nodes. After some time a stable phase develops.

In the limit of a vanishing external field we expect %to %clearly see the
 nucleation %processes 
 determining the growth of a certain phase. 
 Nucleation is exemplified
 %We illustrate nucleation such a process 
 for a small value of $p=0.05$ in Fig. \ref{fig:spreading}. %In 
 The left side of Fig. \ref{fig:spreading} (left)
 % one sees 
 displays
 the initially occurring spreading seeds due to the spontaneous infection dynamics. 
% Later on
 Eventually, contact dynamics (external failure) leads to a local spread of the failure and larger clusters form as illustrated in Fig. \ref{fig:spreading} (right). A video %version 
 of the latter example can be found here: \href{https://vimeo.com/163988456}{video 3 (vanishing spontaneous infection)} \cite{video3}.

\subsection{Phase diagrams and transitions for embedded systems}
\label{sec:pd_trans}
\begin{figure}
\begin{minipage}{0.49\textwidth}
\centering
\includegraphics[width=\textwidth]{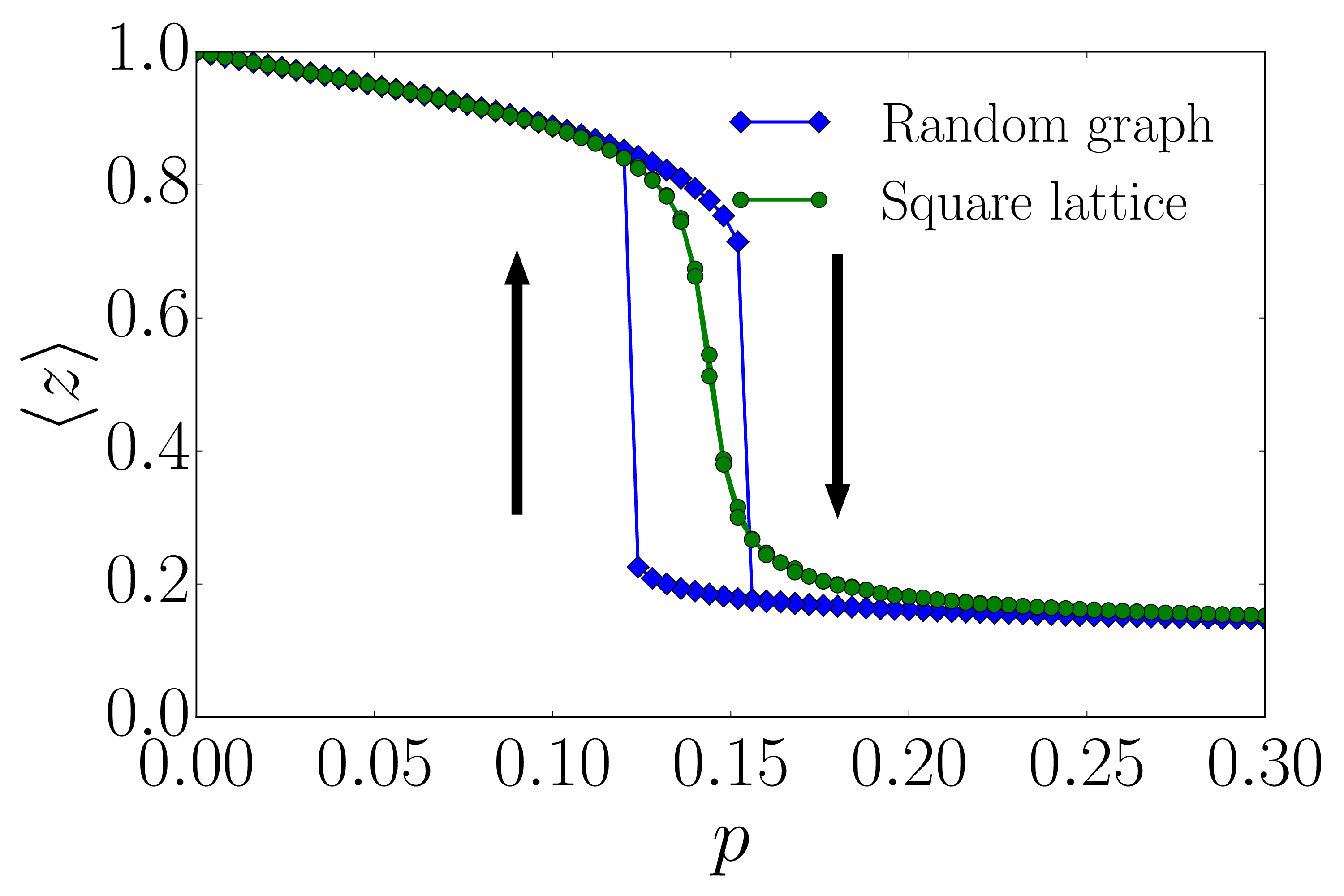}
\end{minipage}
\begin{minipage}{0.49\textwidth}
\centering
\includegraphics[width=\textwidth]{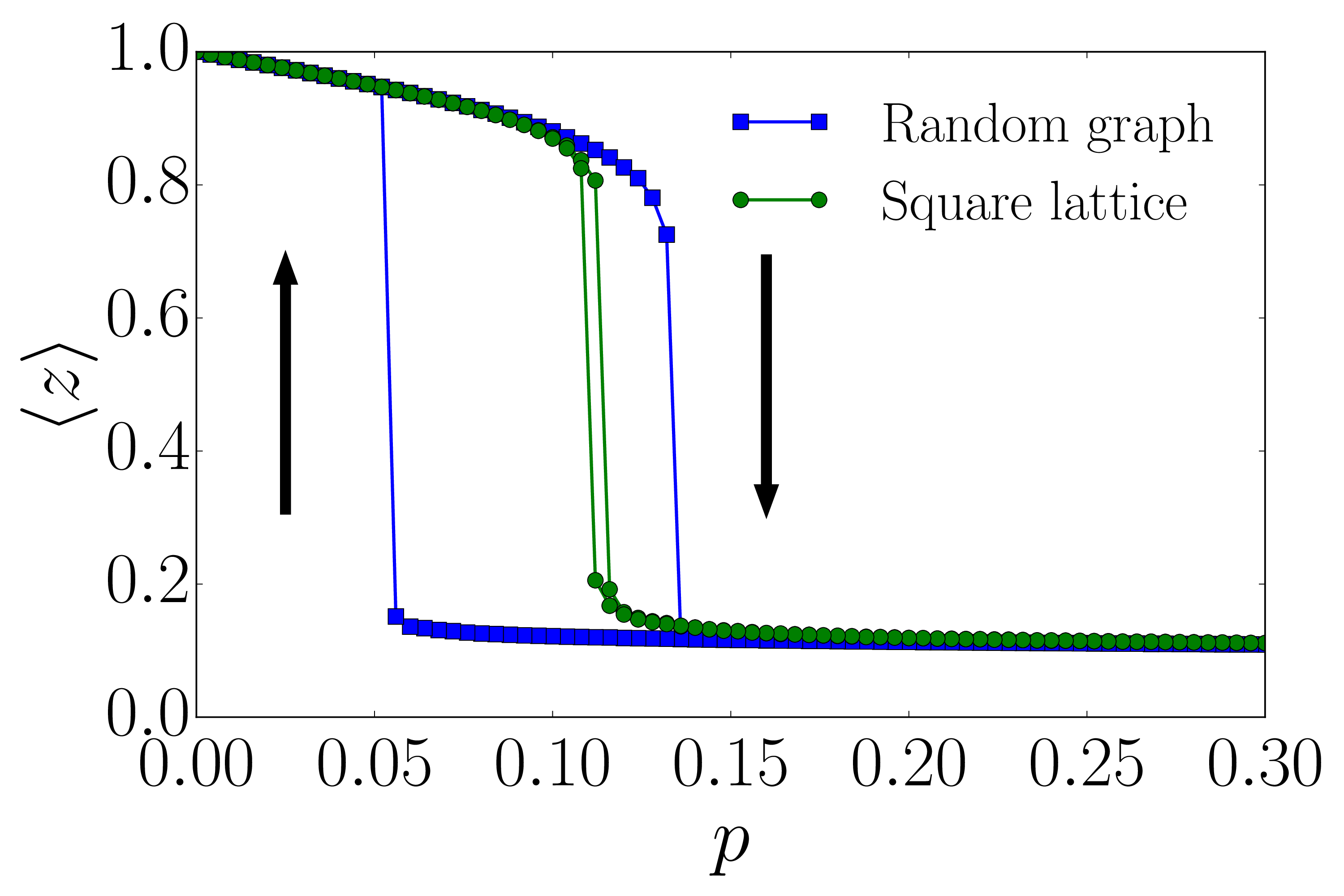}
\end{minipage}
  \caption{
  \textbf{Illustration of hysteresis effects}. % versus continuous transition.} %. %for %Different transitions in a
   %a random graph and a square lattice.} 
   Simulation of the failure-recovery model ($\langle z \rangle$ average of the fraction of active nodes) with $q=1.0$, $q'=0.1$, $m=1$ for a square lattice and a regular random graph both with $k=4$ and $N=512\times 512$ nodes. (left) The transition is discontinuous for $r=0.7$ for the random graph but continuous for the square lattice. (right) Only for large $r/q'$ (shown for $r=1.0$) the transition is discontinuous for both the random graph and the square lattice. The black arrows indicate the direction of the simulation loop.} 
 \label{fig:comp_sl_rg}
\end{figure}
%
%
%
%In mean-field %phase space in Fig. \ref{fig:switching} (right) shows 
%a substantially larger hysteresis region compared to the one of the square lattice is observed. This result shows that the square lattice can be better %controlled against cascading failures occurring in the hysteresis region which in this case are only harmful under large spreading rates. 
%In order to demonstrate that 
Critical failure-recovery dynamics  necessarily occurs close to the hysteresis region.
%On
%The square lattice can be better controlled against cascading failures occurring close to % in 
%the hysteresis region which in this case are only harmful under large spreading rates. In order to demonstrate this
%We illustrate this difference in Fig. \ref{fig:comp_sl_rg}, where 
We study the critical transitions for fixed $r$ and varying $p$ for a regular random graph with degree $k=4$ and for a square lattice. One observes that  for $r=0.7$  the square lattice shows a continuous transition %for $r=0.7$
 whereas %and 
the random graph exhibits a discontinuous transition (Fig. \ref{fig:comp_sl_rg} (left)). 
Since in real systems control parameters often can be only determined approximately, 
this demonstrates that critical failure-recovery dynamics on the lattice can be better controlled compared on a random graph.
%For larger values of $r$, 
When both paths cross the hysteresis region, e.g., for $r/q^\prime=10$, both dynamics show a
 %we also see 
 %a 
 discontinuous transition %in the square lattice 
 (Fig. \ref{fig:comp_sl_rg} (right)). This however requires parameter tuning, that is, large values of the external spreading rate $r/q'$.
\begin{figure}
\begin{minipage}{0.49\textwidth}
\centering
\includegraphics[width=\textwidth]{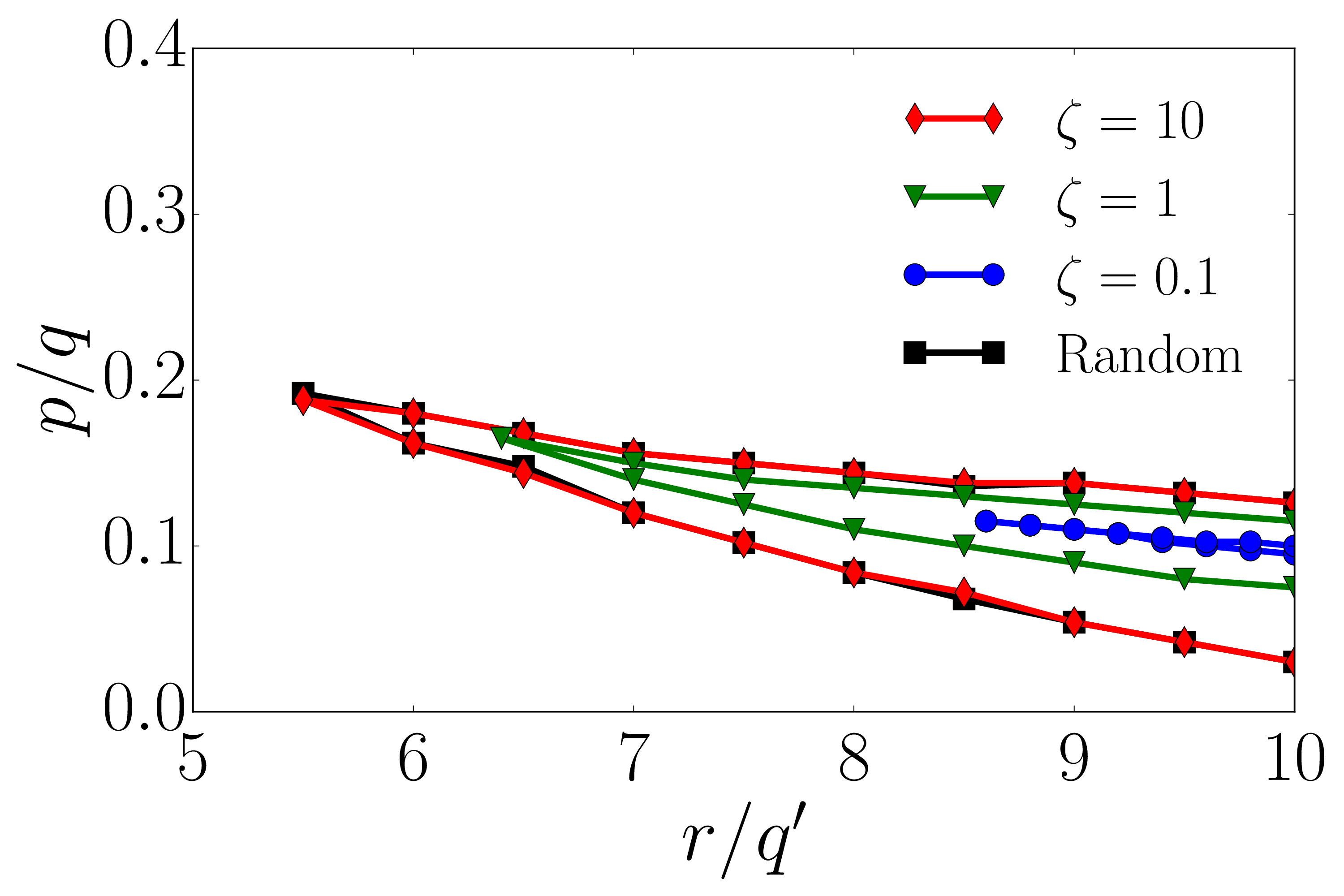}
\end{minipage}
  \caption{
  \textbf{Phase diagrams of %general 
  spatially embedded
  %embedded lattice
  networks
   with $k=4$ in comparison to a random graph.}
  We set $q=1.0$, $q'=0.1$, $m=1$ and perform the simulation for networks with degree $k=4$ and $N=250,000$ nodes. The phase space of a regular random graph with $k=4$ (black squares) is compared to an embedded network 
 % general embedded system 
  with the same degree $k=4$ for different values of the characteristic link length $\zeta$ (blue dots, green inverted triangles, red diamonds). For large $\zeta$ we obtain the phase space of the regular random graph and for small $\zeta$ the square lattice behavior is recovered.} 
 \label{fig:comp_sl_rg_2}
\end{figure}

In order to better understand the dramatic differences between random networks and embedded lattices, we analyze here the transition from a square lattice to a regular random graph. %to the square lattice, 
To this end, we follow the transition model of Danziger et al. \cite{danziger15} and study the phase space of an embedded system with degree $k=4$ where randomly chosen nearest-neighbor links are replaced by longer-range links. The lengths $l$ of the links are distributed according to 
%and 
an exponentially decaying distribution $P(l)\sim \exp\left(-l/\zeta\right)$, with a % of the
 link length $l$ and %decaying according to a 
 characteristic link length 
 $\zeta$. In the thermodynamic limit, %we expect to recover 
 a square lattice in the limit of $\zeta\rightarrow 0$ is recovered, whereas
  in the limit of $\zeta\rightarrow \infty$ we obtain a regular random graph (as all link lengths are equally likely). 
  %We first show 
  %The phase space of a regular random graph with $k=4$,  in comparison to the mean-field behavior is shown in Fig. \ref{fig:comp_sl_rg_2}. 
  %Due to the small value of $k$ the mean-field spinodals are not overlapping with the simulation curves. 
  %However, as shown before in Fig. \ref{fig:phase_spaces_sl_rand}, the \lb{random regular} networks with larger degree are better described by the mean-field solutions.  %We now focus on 
  The phase diagrams of an embedded system with $k=4$ and exponential link length distribution, in the presence of  processes (i-iii), are shown in % illustrated in 
  Fig. \ref{fig:comp_sl_rg_2}. We clearly %see 
  observe the transition from a situation similar to the square lattice for $\zeta=0.1$ to a regular random graph for $\zeta=10$. This again illustrates the strong dependence of the extent of the metastable region on the topology. In other words, a variation in the characteristic link length $\zeta$ causes a very narrow metastable domain ($\zeta=0.1$) to expand into a substantially larger region ($\zeta=10$). Therefore, the results presented in this section have implications for the understanding of the predictability of networks.
\subsection{Oscillatory behavior}
\label{sec:oscillations}
We will briefly describe the possibility of encountering limit cycles in our dynamics. 
We %One possible way of
 investigate %ing
  this behavior by studying %is the analysis of 
  the Lyaponuv function \cite{strogatz14}. In our case, the Lyapunov function $V(a,u_{int})$ is derived from the following equations ($\alpha,\beta > 0$):
\begin{align}
& \frac{d a (t)}{d t} =  -\alpha \frac{\partial V(a,u_{int})}{\partial a}, \\
& \frac{d u_{int} (t)}{d t} =  -\beta \frac{\partial V(a,u_{int})}{\partial u_{int}},
\end{align} 
which are equivalent to
\begin{align}
& \frac{d a (t)}{d t} =  r \sum_k f_k E_k \left(1-a(t)\right)+p \left(1-a(t)\right)- q u_{int} (t) - q' \left(a- u_{int} (t)\right), \\
& \frac{d u_{int} (t)}{d t} = p \left(1-a(t)\right) - q u_{int}(t).
\end{align} 
Without loss of generality, we set $\beta=1$ and compute $V\left(a,u_{int}\right)$ to:
\begin{align}
\begin{split}
&V \left(a,u_{int}\right)=q \frac{u_{int}^2}{2}-p u_{int}\left(1-a \right)- \frac{p}{q-q'} \left[p \left(a-\frac{a^2}{2}\right)-q' \frac{a^2}{2} +\right. \\
&r \left. \sum_k f_k \sum_{j=0}^m \sum_{l=0}^{j+1}\binom {k} {k-j} \binom {j+1} {l} (-1)^l \frac{a^{k-j+l+1}}{k-j+l+1}  \right],
\end{split}
\end{align}
where we used the binomial theorem and set $\alpha=\left(q-q'\right)/p$. We find that
\begin{align}
\begin{split}
& \frac{d V\left(a,u_{int}\right)}{d t}=\frac{d a (t)}{d t} \frac{ \partial
V\left(a,u_{int}\right)}{\partial a}+\frac{d u_{int}(t)}{ d t}\frac{ \partial V\left(a,u_{int}\right)}{\partial u_{int}} \\
&=-\alpha \left(\frac{ \partial V\left(a,u_{int}\right)}{\partial a}\right)^2-\beta \left(\frac{ \partial V\left(a,u_{int}\right)}{\partial u_{int}}\right)^2 < 0,
\end{split}
\end{align}
if $q>q'$ since $\beta=1$.
\begin{figure}
\begin{minipage}{0.49\textwidth}
\centering
\includegraphics[width=\textwidth]{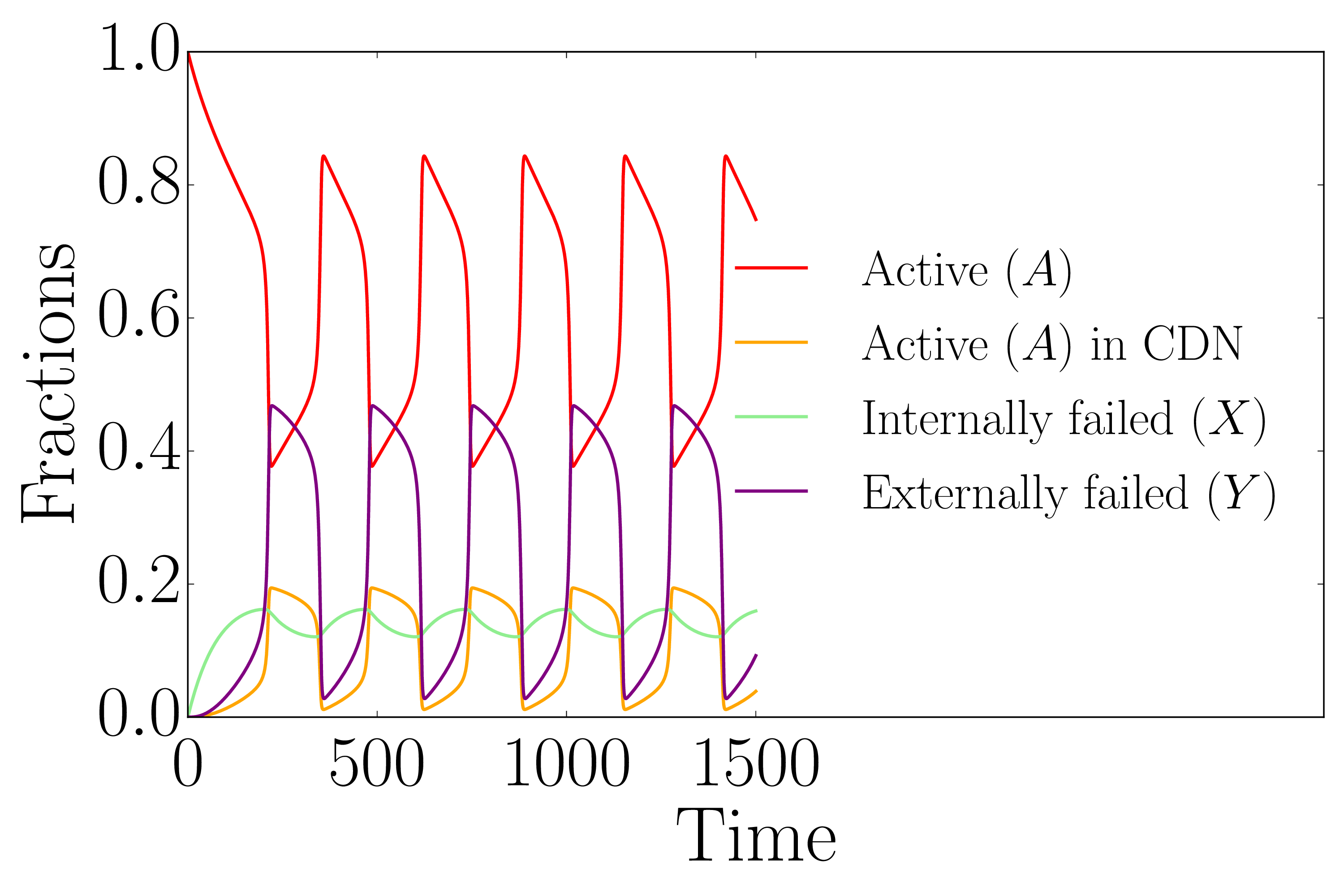}
\end{minipage}
\begin{minipage}{0.49\textwidth}
\centering
\includegraphics[width=\textwidth]{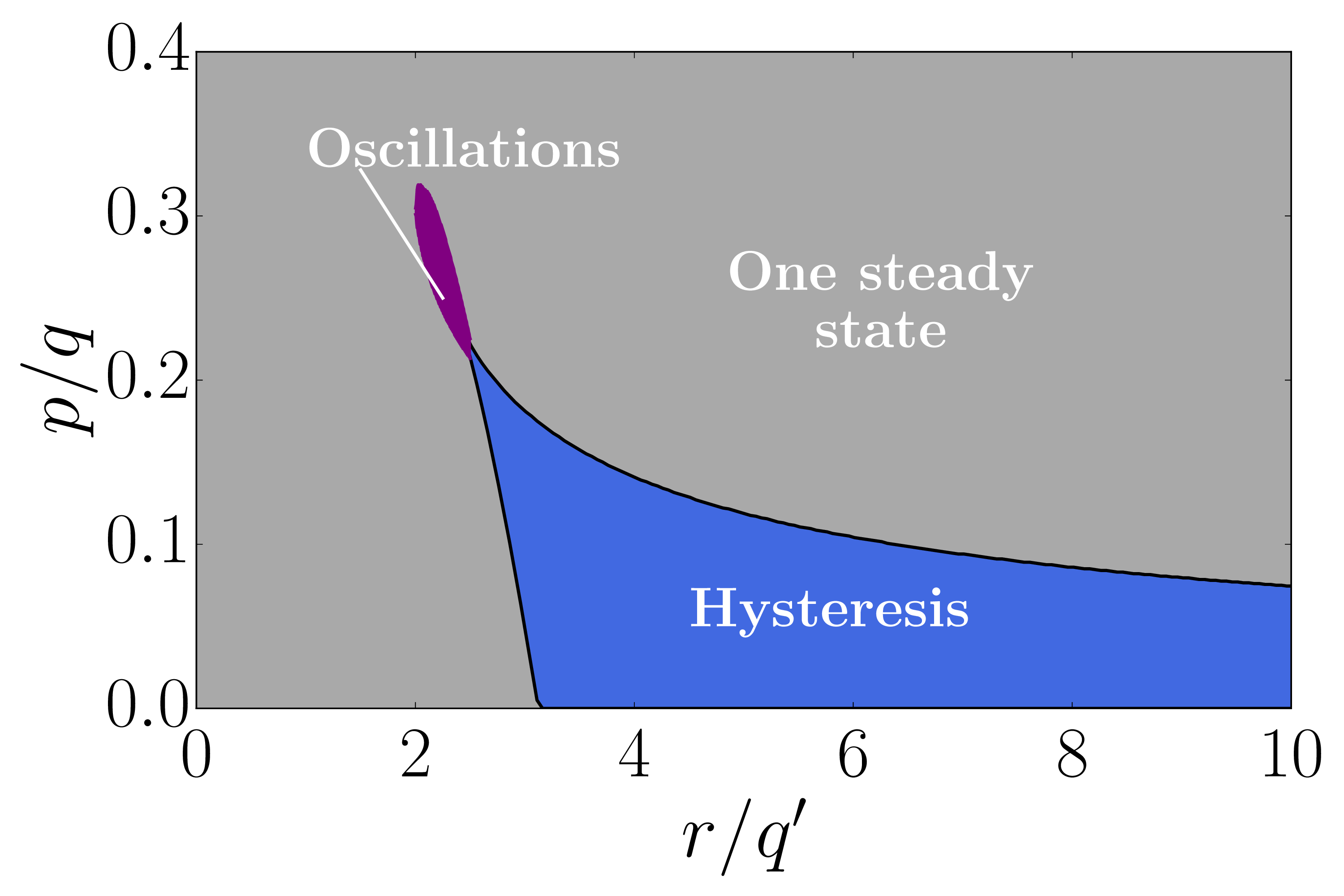}
\end{minipage}
  \caption{\textbf{Oscillatory behavior (limit cycles) for $q' > q$.} 
  (left) We show the mean-field time evolution of different compartments, cf. Fig. \ref{fig:model}, for $p/q=19/81$, $q=0.01$, $r/q'=3125/1296$, $q'=1.0$, $k=4$ and $m=1$. The values of $p$ and $r$ correspond to the ones of the bifurcation point as described in Appendix \ref{sec:critical_behavior}. We clearly see the oscillatory behavior as a consequence of $q' > q$. (right) The phase space for $k=4$, $q=0.01$, $q'=1.0$ and $m=1$ now also displays an oscillatory regime (purple).} 
 \label{fig:oscillations}
\end{figure}
For $q>q'$ we therefore expect no oscillatory behavior, i.e. no closed orbits. However, for $q'>q$, we show in Fig. \ref{fig:oscillations} the existence of closed orbits. The purple region in Fig. \ref{fig:oscillations} (right) illustrates the regime where we measured a periodic orbit analyzing the Fourier transformed time evolution of the fraction of active nodes. We also show the oscillatory dynamics of a non-embedded regular network in \href{https://vimeo.com/172921549}{video 4 (oscillations)} \cite{video4}. As discussed in Appendix \ref{sec:connection}, for $q=q'$, the differential equations Eqs. \eqref{eq:internal_rate} and \eqref{eq:external_rate} can be decoupled and one obtains a single first-order differential equation which has no periodic solutions \cite{strogatz14}. 

This demonstrates that the phase diagram is substantially more complex than previously believed. Specifically, limit cycles occur for $q'/q<1$ in a narrow region in the phase diagram. This deterministic behavior is markedly different from the stochastic switching dynamics in the hysteresis region but likewise challenges control.
\section{Discussion}
We have derived a unifying framework for the
interplay between failure, damage spread and recovery in spatially embedded and random networks.
The theoretical description 
links
diverse phenomena such as 
complex contagion and phase-switching due to metastability and the occurrence of cusp catastrophes.
The number of failed neighbors necessary to allow external failure to act on a node
is a crucial parameter of the system.
%, cascading  and the topology of different embedded networks. 
%In particular, 
Our analysis revealed that the phase space is substantially more complex than previously known
owing to the coexistence of
%In particular, we identified when and how
 limit cycles and random phase switching within hysteresis. % can coexist.

%, the resulting dynamics is very diverse, including hysteresis and limit cycles. 
%
We analytically demonstrated that the mean-field description of the stochastic model systems
is 
%Our model reveals a phase space similar 
equivalent to cusp catastrophes with two bifurcation lines enclosing a metastable domain where two stable stationary states coexist. Inside this metastable region, large fractions of nodes suddenly %, i.e. in an uncontrollable manner, 
fail and recover. % within a short period of time.
We propose %to interpret 
the hysteresis (metastable) area as a
 predictability measure for the state of the system of a given topology and dynamics. 
 Our results show that the transition from a 
 random 
 %non-embedded
  regular network to an %spatially 
  embedded network with a short characteristic link length is characterized by a dramatic shrinking of the metastable domain.
 This % what 
 suggests that embedded systems with short characteristic link lengths whose dynamics is captured by processes (i-iii) % basic 
 are substantially more robust against abrupt spontaneous and cascading failures compared to non-embedded systems.

Moreover, 
we have also shown that our theoretical framework is able to describe essential features of the model's time evolution
and that it captures spontaneous failure as an external field in analogy to magnetic systems.
However, based on the connection to contact process dynamics we find that %don't expect
 the model does not %to 
 belong to the Ising universality class as %stated 
 conjectured earlier \cite{majdandzic14}. The arguments in Appendices \ref{sec:connection} and \ref{sec:critical_behavior} show the similarities to the (non-equilibrium) contact process belonging to the directed percolation universality class \cite{henkel08}. 
 %And, 
 In fact, as %already 
 mentioned by Grassberger
 %in Ref. 
 \cite{grassberger82}, relating this dynamics to the Ising universality class would mean an extension of the universality hypothesis from models with detailed balance to models without it.
%Two types of  
   %The loss of predictability is two-fold.
    Unpredictability observed in the hysteresis region results from two effects.
    For finite systems, unpredictable random phase switching between two unstable states is observed.
   In the thermodynamic limit, however, small fluctuations in the initial phase of the systems dynamics determine 
   the stationary (stable) state of the system.
 
 Our framework helps 
%a first attempt
 to better understand predictability and controllability in spatially embedded and random systems 
%and, in addition, unifies pre
where spontaneous recovery, % can mitigate %or even counterbalance 
spontaneous failure and cascading failure lead to a remarkably complex dynamic interplay.

\acknowledgments
We acknowledge financial support from the ETH Risk Center (grant number RC SP 08-15) 
 and ERC Advanced grant number FP7-319968 FlowCCS of the European Research Council. SH acknowledges the MULTIPLEX (No. 317532) EU
project, the Israel Science Foundation, the Italian-Israel and Japan-Israel Most, ONR and DTRA for financial support. We are very thankful to Michael M\"{a}s for his thoughtful comments on complex contagion phenomena. We thank Linda Mathez for assisting in the preparation of the figures in sections \ref{sec:switching} and \ref{sec:pd_trans}.
\section*{Author contribution statement}
L.B. carried out computational simulations, analytical calculations and wrote the first draft of the manuscript. All authors contributed equally to the ideas, the interpretation and the presentation.
\section*{Additional information}
\textbf{Competing financial interests} The author(s) declare no competing financial interests.
\clearpage
\appendix
\section{Connection to other models}
\label{sec:connection}

\begin{figure}
\begin{minipage}{0.49\textwidth}
\centering
\includegraphics[width=\textwidth]{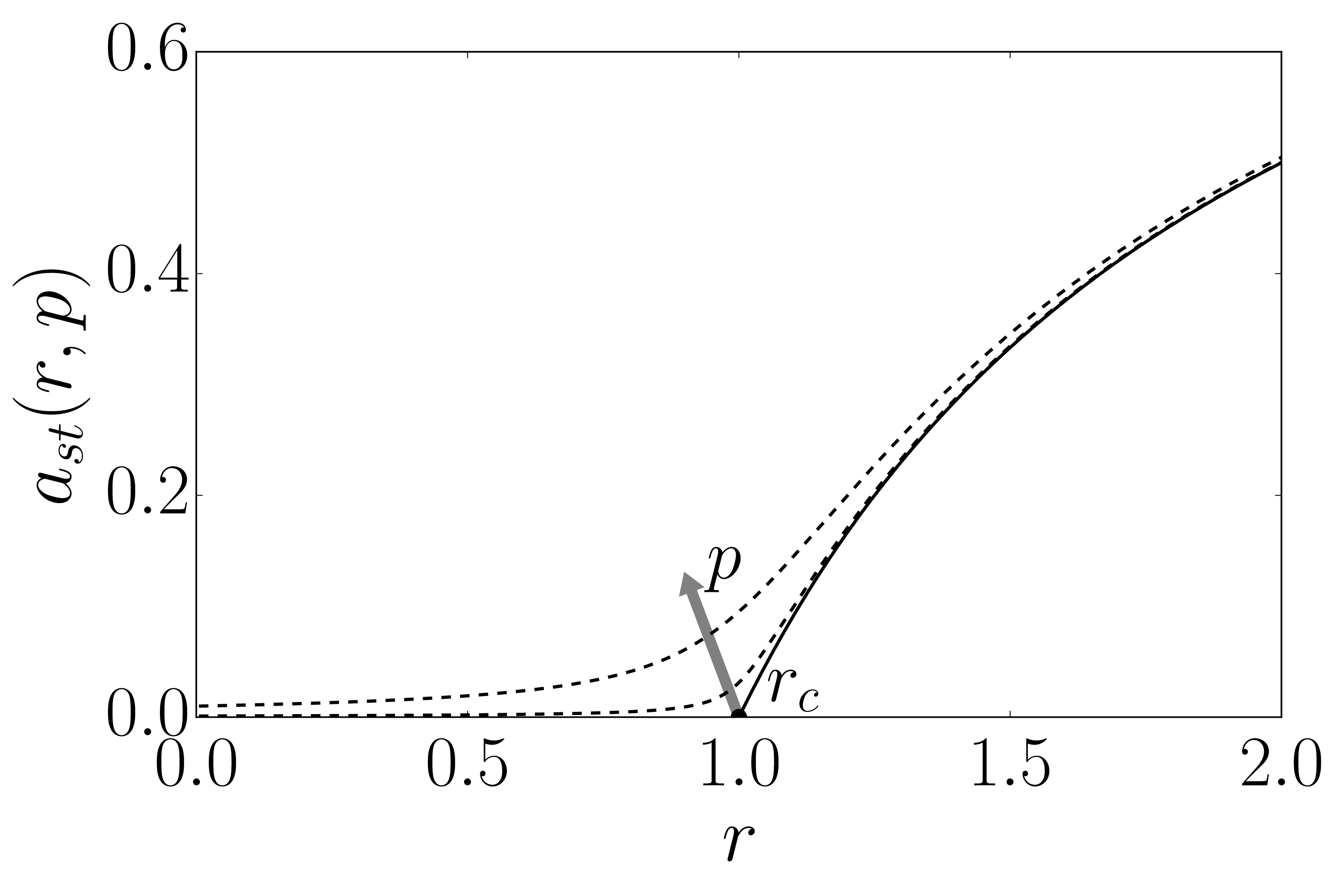}
\end{minipage}
\begin{minipage}{0.49\textwidth}
\centering
\includegraphics[width=\textwidth]{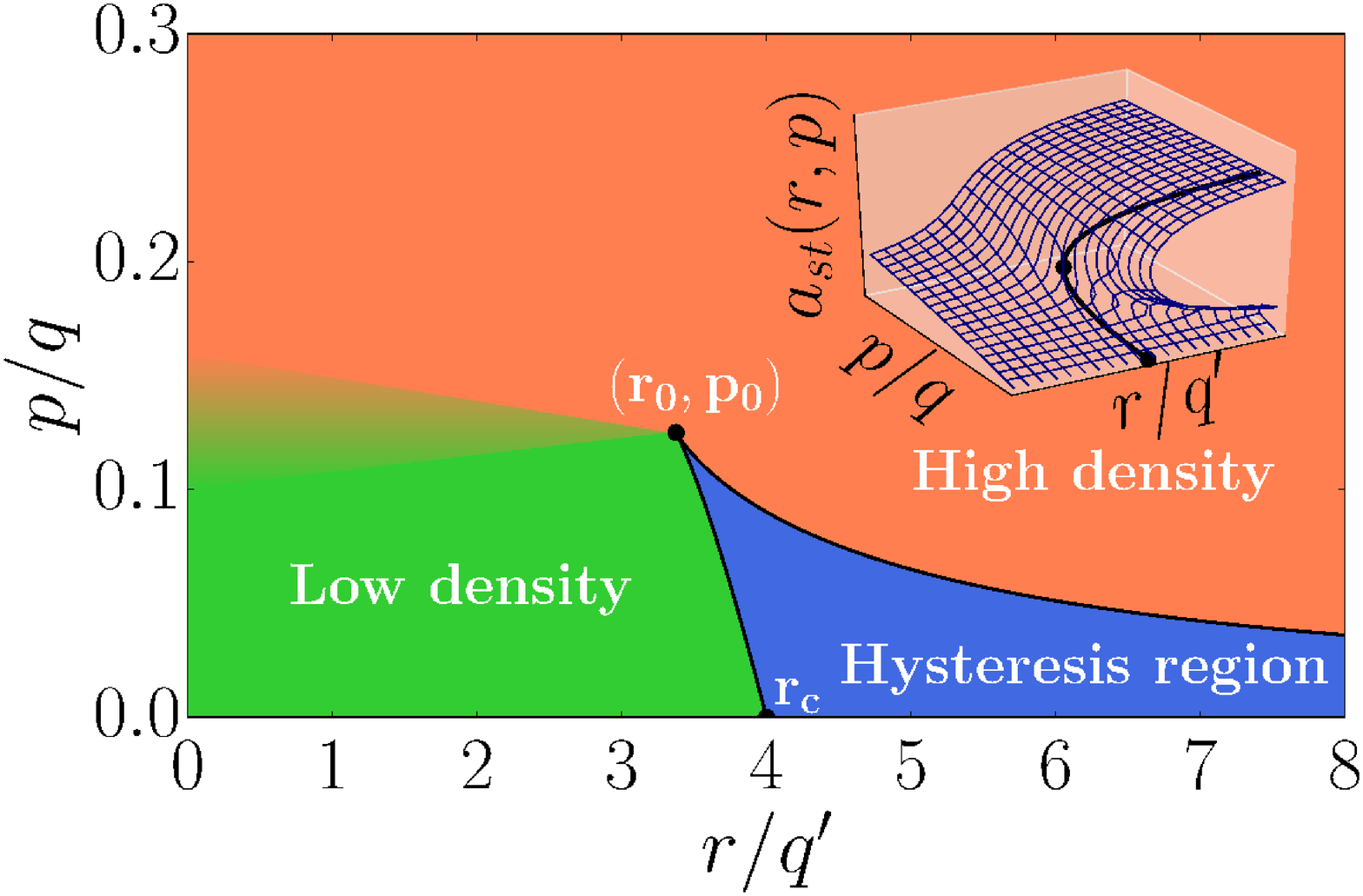}
\end{minipage}
  \caption{\textbf{Analogy to the contact process and Schl\"ogl's second model.} (left) The order parameter $a_{st}(r)$ as a function of the external failure rate $r$ for $k=1$ and $m=0$ (mean-field). Analogous to the contact process the black solid line corresponds to the situation where the vanishing spontaneous infection term $p\rightarrow 0$ leads to a second-order phase transition at $r=r_c$. The dashed lines show a smeared out transition due to the non-zero spontaneous infection rates $p=0.001, 0.01$ \cite{marro05}. (right) The phase space for $k=2$ and $m=0$ (mean-field). One clearly sees the hysteresis region where two states coexist (low density and high density failure phases). The spinodals (black solid lines) merge at the bifurcation point $(r_0,p_0)=(27/8,1/8)$. The critical point, $r_c=4$ indicates the transition point without additional field-like term ($p=0$). This situation is similar to Schl\"ogl's second model \cite{grassberger82}, the phase space of cusp catastrophes or imperfect bifurcations \cite{strogatz14}.} 
 \label{fig:cp_field}
\end{figure}

To draw a connection to other models we first simplify the two coupled rate equations Eqs. \eqref{eq:internal_rate} and \eqref{eq:external_rate}. We therefore set $q=q'=1$ (excluding limit cycles, cf. Sec. \ref{sec:oscillations}) and add Eqs. \eqref{eq:internal_rate} and \eqref{eq:external_rate} to obtain:
\begin{equation}
\frac{d a (t)}{d t}= r \sum_k f_k E_k \left(1-a(t)\right)+p \left(1-a(t)\right)- a(t).
\label{eq:cp_spont_gen}
\end{equation}
In the limit of a regular degree distribution with $k=1$ and $m=0$, and under the assumption of dynamically rewiring links between nodes, we find exact correspondence to the contact process dynamics with spontaneous infection \cite{marro05}:
\begin{equation}
\frac{d a (t)}{d t}= r a(t) \left(1-a(t)\right)+p \left(1-a(t)\right)- a(t).
\label{eq:cp_spont}
\end{equation}
The latter equation describes nothing but contact process dynamics with a smeared out second order phase transition transition due to the additional spontaneous infection term. We illustrate the stationary state $a_{st}(r)$ (order parameter) as a function of the external failure rate $r$ in Fig. \ref{fig:cp_field} (left). In the limit of vanishing spontaneous failure $p\rightarrow 0$ one encounters a second order phase transition. At the critical point $r_c=1$ the order parameter grows as $a_{st}(r)\propto (r-r_c)^\beta$ with $\beta=1$. A non-zero spontaneous failure term leads to a smeared out transition. This situation is similar to the one in the Ising model with an applied magnetic field which also removes the second order phase transition. However, unlike in the Ising model the field equivalent satisfies the condition $p>0$ and we are restricted to one of the two roots defining the stationary state of Eq. \eqref{eq:cp_spont}:
\begin{equation}
a_{st}(r,p) = \frac{1}{2 r} \left[r-p-1 + \sqrt{(r-p-1)^2 + 4 r p}\right].
\label{eq:cp_spont_st}
\end{equation}
Close to the critical point $r_c=1$, i.e. $r \rightarrow r_c$, we find $a_{st}(r_c,p) \propto p^{1/\delta_h}$ with the field exponent $\delta_h=2$ in the mean-field situation.

In order to see the influence of the coupling parameter $m$ on the dynamics, we now turn towards the case $k=2$ and are free to set $m=0, 1, 2$. For $m=2$ all neighborhoods are critically damaged by definition and the stationary state is given by $a_{st}(r,p)=(r+p)/(1+r+p)$. In particular, this solution is obtained for all regular graphs with degree $k$ and $m=k$ since $E_k=\sum_{j=0}^{k}  \binom {k} {k-j} a^{k-j} (1-a)^j =1$.

The situation is different for $m=1$ where at least one neighbor of a given node needs to fail in order to allow external failure acting on the node. This is again in accordance with the contact process where also at least one failed neighbor is necessary to turn on the spreading dynamics. We also find the corresponding exponents $\beta=1$ and $\delta_h=2$ in the vicinity of $r_c=1/2$. In general, we expect this behavior for any regular graph with degree $k$ and $m=k-1$ since $\lim_{a\rightarrow 0} E_k=\sum_{j=0}^{k-1}  \binom {k} {k-j} a^{k-j} (1-a)^j = k a + \mathcal{O}(a^2)$ (at the critical point). That is the reason why we again find the contact process exponents in the latter example and a critical value of $r_c=1/2$ which is just the critical point of Eq.  \eqref{eq:cp_spont} divided by $k$.

Another interesting behavior is found for $m=0$. Without spontaneous failure term, the rate equation describes a pair-creation contact process \cite{tome15} and taking this term into account yields a variant of Schl\"ogl's second model \cite{grassberger82, vellela09}. Setting $p=0$, the stationary state for $r>r_c=4$ is given by $a_{st}(r)=1/2 (1 + \sqrt{1-4/r})$ and $a_{st}(r)=0$ for $r<r_c$. The phase diagram for $m=0$ and $p\geq 0$ is illustrated in Fig. \ref{fig:cp_field} (right). Two spinodals define the hysteresis region where two states coexist. As for cusp catastrophes \cite{strogatz14}, this region is the projection of the hysteresis set from three dimensions into plane space, cf. inset in Fig. \ref{fig:cp_field} (right). In this example, the spinodals are defined by $\Delta = 0$ where the discriminant $\Delta=-r (4 + 4 p^3 - r + 4 p^2 (3 + 2 r) + 4 p (3 - 5 r + r^2))$. For $\Delta < 0$ two stable coexisting steady states exist while for $\Delta > 0$ there is only one. The spinodals merge at the bifurcation point (cusp point) characterized by $(r_{0}, p_{0})=(27/8,1/8)$ where $\lim_{(r,p)\rightarrow (r_0,p_0)} \partial r/\partial a = 0$. At the bifurcation point the quantity $\Delta a_{st} (r,p_0)=a_{st}(r,p_0)-a_{st}(r_0,p_0)$ increases with $r$ as $\Delta a_{st} (r,p_0)\propto (r-r_0)^{\tilde{\beta}}$ and with $p$ as $\Delta a_{st} (r_0,p)=a_{st}(r_0,p)-a_{st}(r_0,p_0)\propto (p-p_0)^{1/\tilde{\delta_h}}$ where $a_{st}(r_0,p_0)=1/3$, $\tilde{\beta}=1/3$ and $\tilde{\delta_h}=3$. For this example, it is straightforward to show that the quadratic term in the Taylor expansion of $f(a_{st},r,p)=r a_{st}^2(1-a_{st})+p(1-a_{st})-a_{st}$ around $a_{st}(r_0,p_0)=1/3$ vanishes yielding the characteristic polynomial of the cusp catastrophe \cite{zeeman79}. In general, the spontaneous recovery model resembles the dynamics of a modified contact process where a certain minimum number of nodes is necessary to turn on the spreading dynamics \cite{tome15}. As already mentioned in previous studies and as discussed in the latter examples, slight modifications of the standard contact process dynamics might have dramatic effects on the system's dynamics leading to uncontrollable abrupt transitions \cite{boettcher14, boettcher16}. 
\section{Critical behavior on the square lattice}
\label{sec:critical_behavior}
\begin{figure}
\begin{minipage}{0.49\textwidth}
\centering
\includegraphics[width=\textwidth]{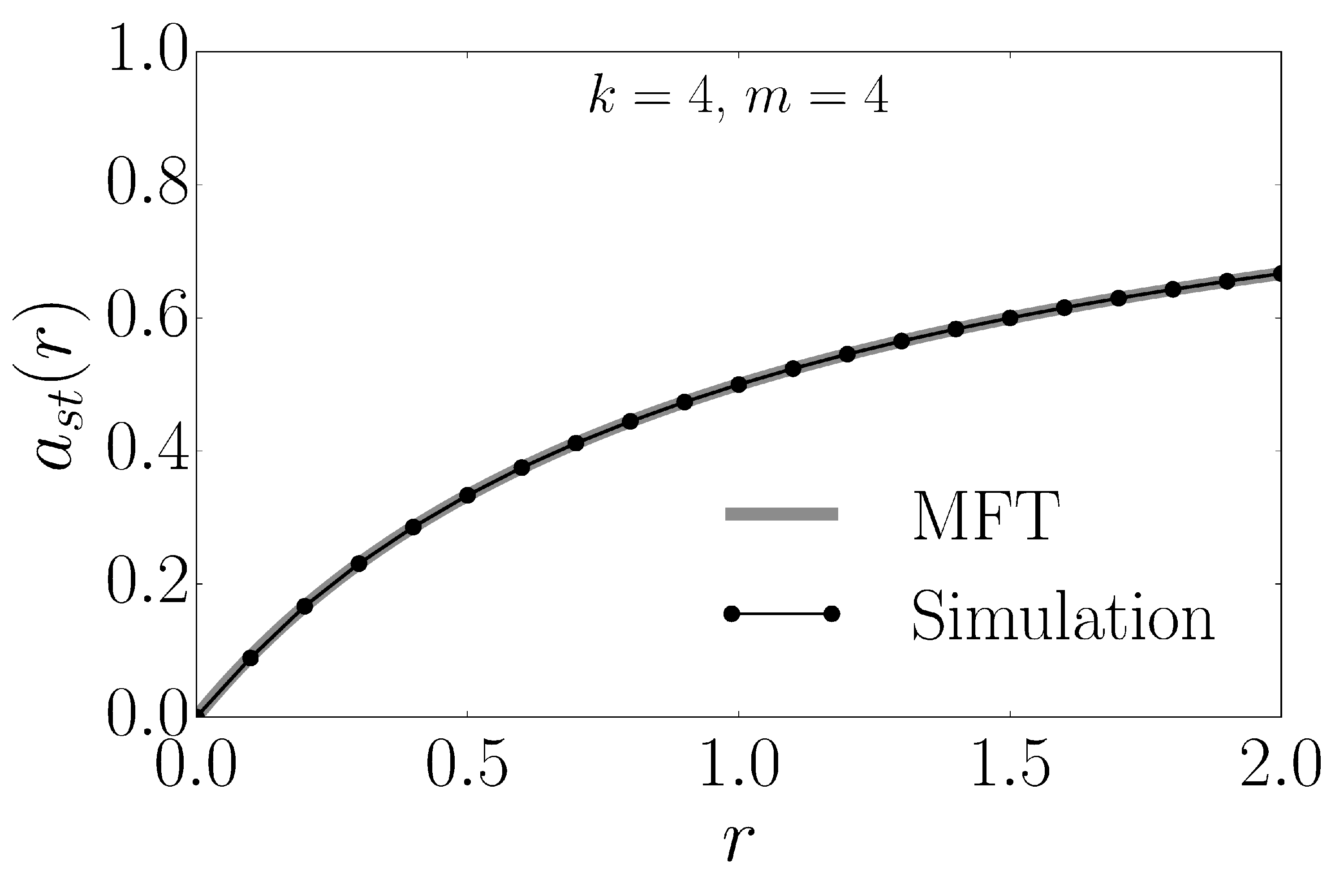}
\end{minipage}
\begin{minipage}{0.49\textwidth}
\centering
\includegraphics[width=\textwidth]{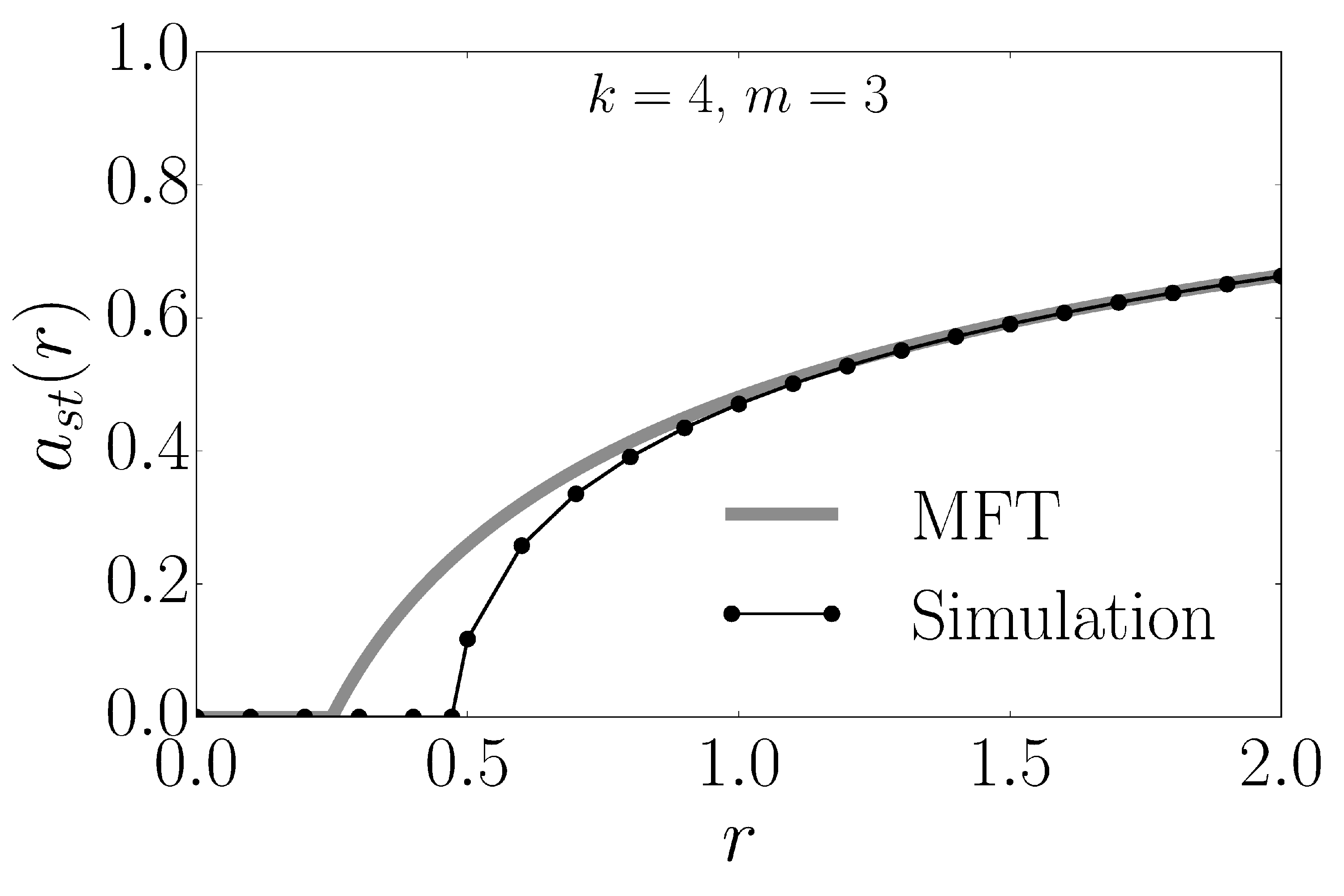}
\end{minipage}
  \caption{\textbf{Growth of the order parameter on a square lattice for different $m$.} Simulation of the spontaneous recovery model without internal failure dynamics ($p=0$) and $q'=1.0$ for different $m$.  (left) The order parameter $a_{st}(r)$ as function of $r$ for $m=4$ and (right) for $m=3$. The simulations have been performed on a square lattice with $N=1024\times 1024$ nodes.} 
 \label{fig:no_field_sl_m34}
\end{figure}

We will now study the critical behavior of the dynamics in a system with degree $k=4$ since there are four nearest-neighbors for every node on a square lattice. Consequently, we have five possibilities of choosing $m=0,1,2,3,4$.

We start with the case $m=4$ for which a CDN even exists when there is no failed neighboring lattice site, i.e. external failure acts all the time independent of the nearest-neighbors' state, cf. Appendix \ref{sec:connection}. Setting $q'=1.0$, the MFT yields for the stationary state of failed nodes $a_{st}(r)=r/(1+r)$ (without field-like spontaneous failure). As long as $r>0$ we find a non-zero fraction of failed nodes in the network. We see in Fig. \ref{fig:no_field_sl_m34} (left) that the results obtained through simulations on a square lattice are well described by the MFT. An additional field-like contribution of the spontaneous failure $p$ and $q=1.0$ yields $a_{st}(r)=(r+p)/(1+r+p)$, cf. Appendix \ref{sec:connection}.
\begin{figure}
\begin{minipage}{0.49\textwidth}
\centering
\includegraphics[width=\textwidth]{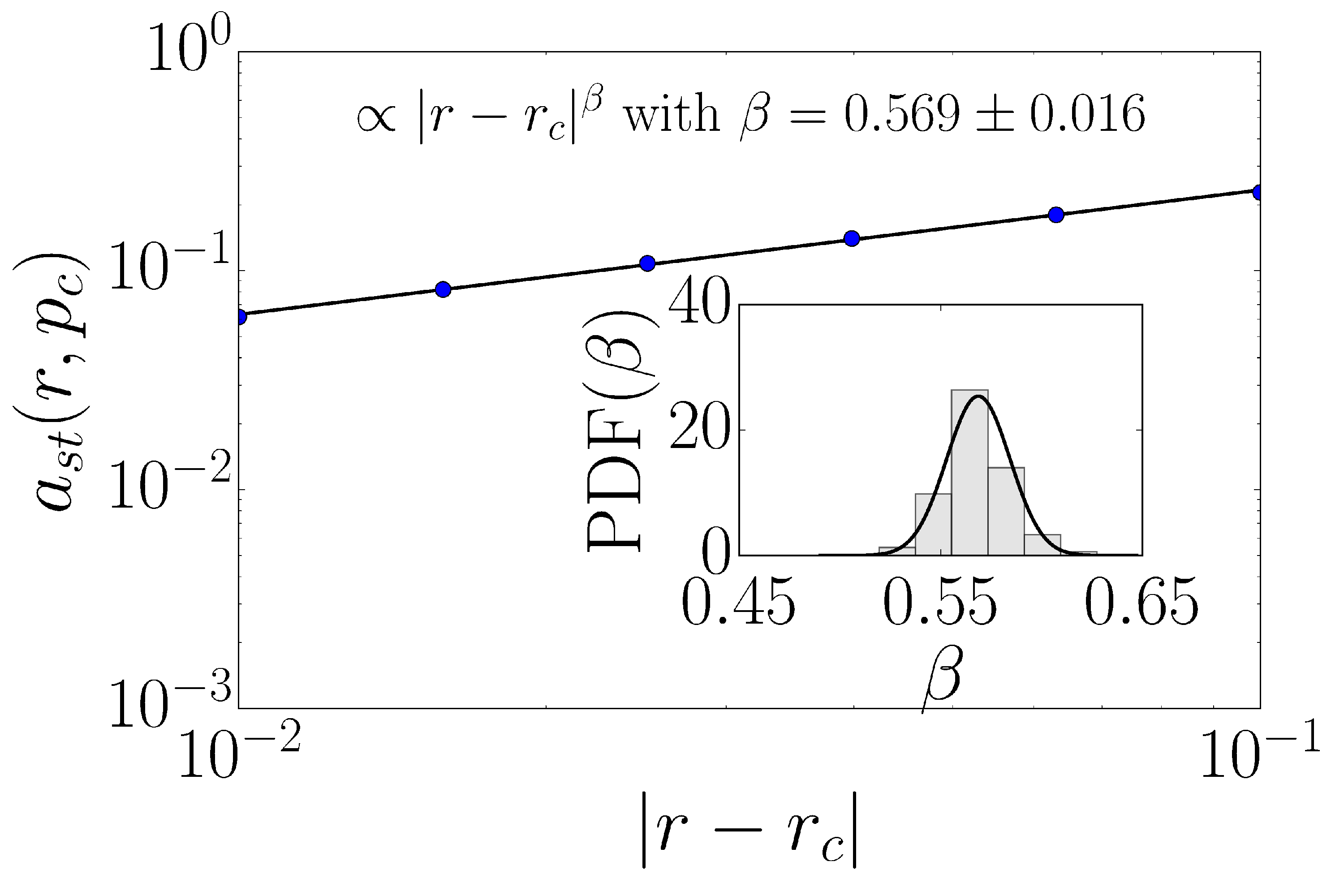}
\end{minipage}
\begin{minipage}{0.49\textwidth}
\centering
\includegraphics[width=\textwidth]{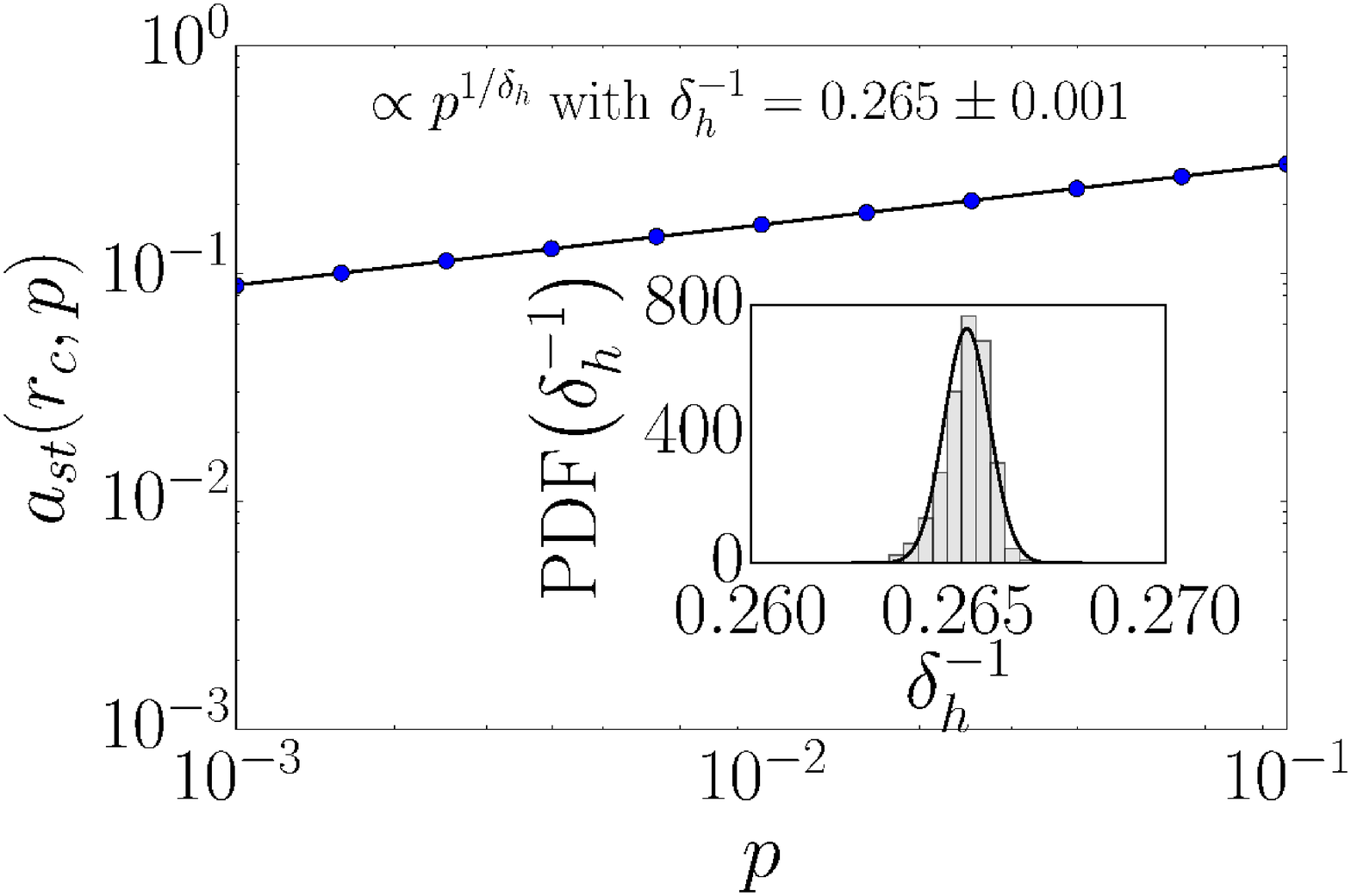}
\end{minipage}
  \caption{\textbf{Critical exponents of the square lattice with $m=3$.} Simulation of the spontaneous recovery model with $p=0$, $q'=1.0$ and $m=3$.  (left) The order parameter $a_{st}$ in the vicinity of the critical point $r_c=0.47(1)$ for different $r$. The exponent found indicates contact process dynamics where $\beta=0.586(14)$ \cite{Moreira96}. (right) The order parameter $a_{st}$ at the critical point $r_c=0.47(1)$ for small values of $p$. The critical exponent measured also indicates contact process dynamics with $ \delta_h^{-1}=0.285(35)$  \cite{adler87}. The simulations have been performed on a square lattice with $N=1024\times 1024$ nodes ($1500$ samples). The insets show the PDF's of the exponent's bootstrap analysis.} 
 \label{fig:no_field_m3_beta_deltah}
\end{figure}

For $m=3$ we expect to find dynamics analogous to the contact process, since only one failed neighbor is needed to let the neighboring nodes fail. This has been described in Appendix \ref{sec:connection} and a non-zero stationary state $a_{st}(r)$ is found if $r>r_c$ ($r_c=1/4$ MFT). From MFT we also find $a_{st}(r)\propto (r-r_c)^\beta$ with $\beta=1$. At $r_c$ the order parameter grows continuously. Applying the field term in this example one finds $a_{st}(r_c,p) \propto p^{1/\delta_h}$ with $\delta_h=2$. We show the order parameter $a_{st}(r)$ as a function of $r$ for the square lattice in comparison with MFT in Fig. \ref{fig:no_field_sl_m34} (right). We also analyzed the critical behavior in the vicinity of the critical point $r_c=0.47(1)$ of the square lattice (see Fig. \ref{fig:no_field_sl_m34} (right)). The growth of the order parameter with $\beta=0.569(16)$ (Fig. \ref{fig:no_field_m3_beta_deltah} (left)) and $\delta_h^{-1}=0.265(1)$ (Fig. \ref{fig:no_field_m3_beta_deltah} (right)) agrees with the corresponding contact process values $\beta=0.586(14)$ \cite{Moreira96} and $ \delta_h^{-1}=0.285(35)$  \cite{adler87}. We thus conclude that the model resembles standard contact process dynamics in this case.

For $m=2$ the transition in MFT is characterized by a jump at $r_c=1.226$ from zero to $a_{st}(r_c)=0.322$. In the simulations on the square lattice we found strong dependence on the initial conditions. %and we do not discuss all the different steady states here.

We did not find a non-zero value of $a_{st}(r)$ for $m<2$ having a circle-shaped seed as initial condition on the square lattice. The situations where $m=0$ or $1$ mean that three or four failed neighbors are needed to turn on external failure. Starting from a circle-shaped seed the dynamics will never reach a stable configuration besides the absorbing state (all nodes are active).
Nevertheless, we are able to study the dynamics for $m=1$ as before by introducing the field-like spontaneous failure term again ($p>0$). In the mean-field situation Eq. \ref{eq:cp_spont_gen} yields for the bifurcation point $(r_0,p_0)=(3125/1296, 19/81)\approx (2.41, 0.23)$ and $a_{st}(r_0,p_0)=0.4$. This point is also shown in Fig. \ref{fig:phase_spaces_sl_rand} (right). Similar to the arguments in Appendix \ref{sec:connection}, it is again straightforward to show that the quadratic term in the Taylor expansion of the polynomial describing the stationary states around $a_{st}(r_0,p_0)=0.4$ vanishes yielding the characteristic polynomial of the cusp catastrophe \cite{zeeman79}. The black lines in the latter figure characterize the hysteresis region with two coexisting stationary states similar to Fig. \ref{fig:cp_field} (right). From MFT we find $\Delta a_{st} (r,p_0)\propto |r-r_0|^{\tilde{\beta}}$ with $\tilde{\beta}=1/3$ and $\Delta a_{st} (r_0,p)\propto |p-p_0|^{1/\tilde{\delta_h}}$ with $\tilde{\delta_h}=3$.
\begin{figure}
\begin{minipage}{0.49\textwidth}
\centering
\includegraphics[width=\textwidth]{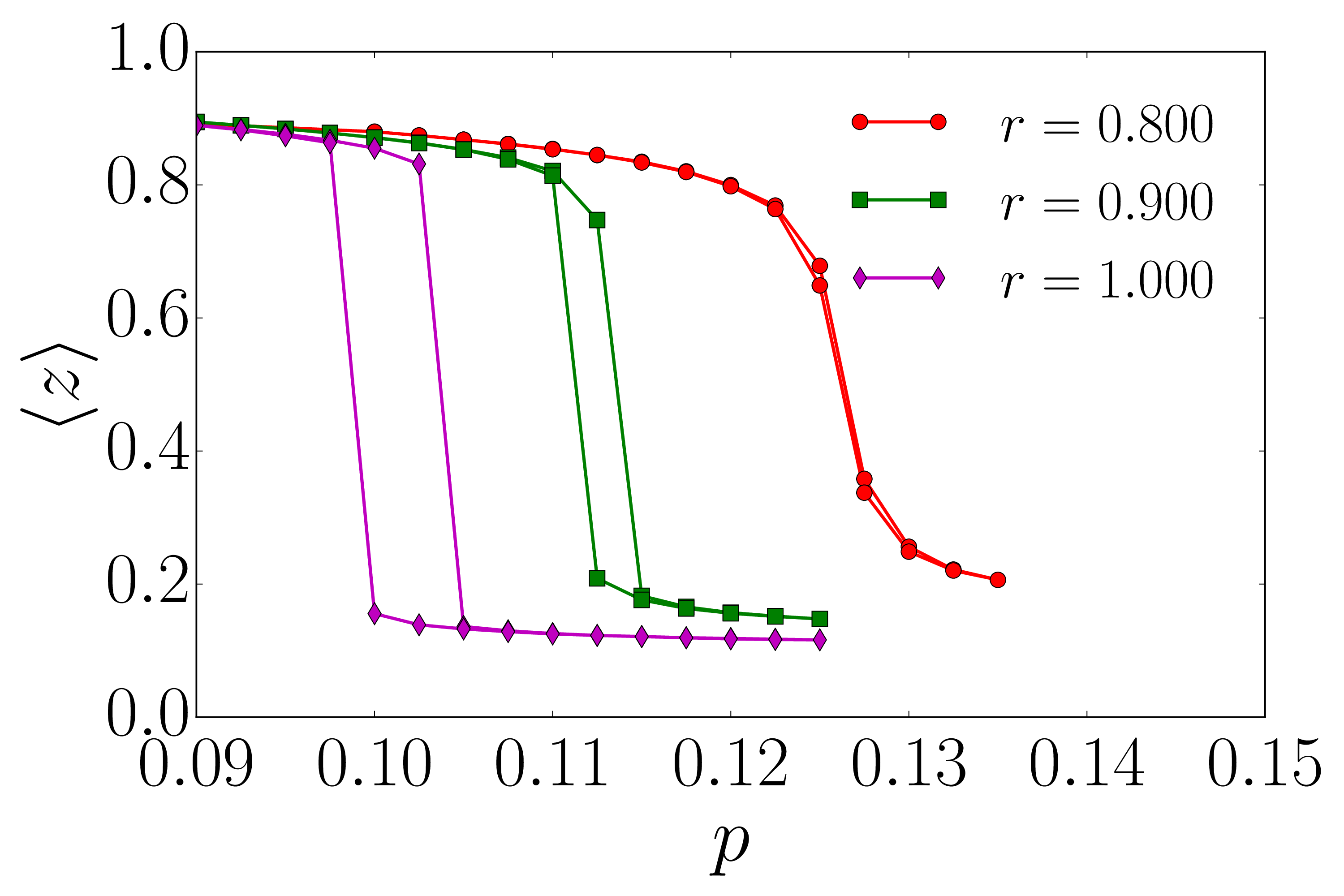}
\end{minipage}
\begin{minipage}{0.49\textwidth}
\centering
\includegraphics[width=\textwidth]{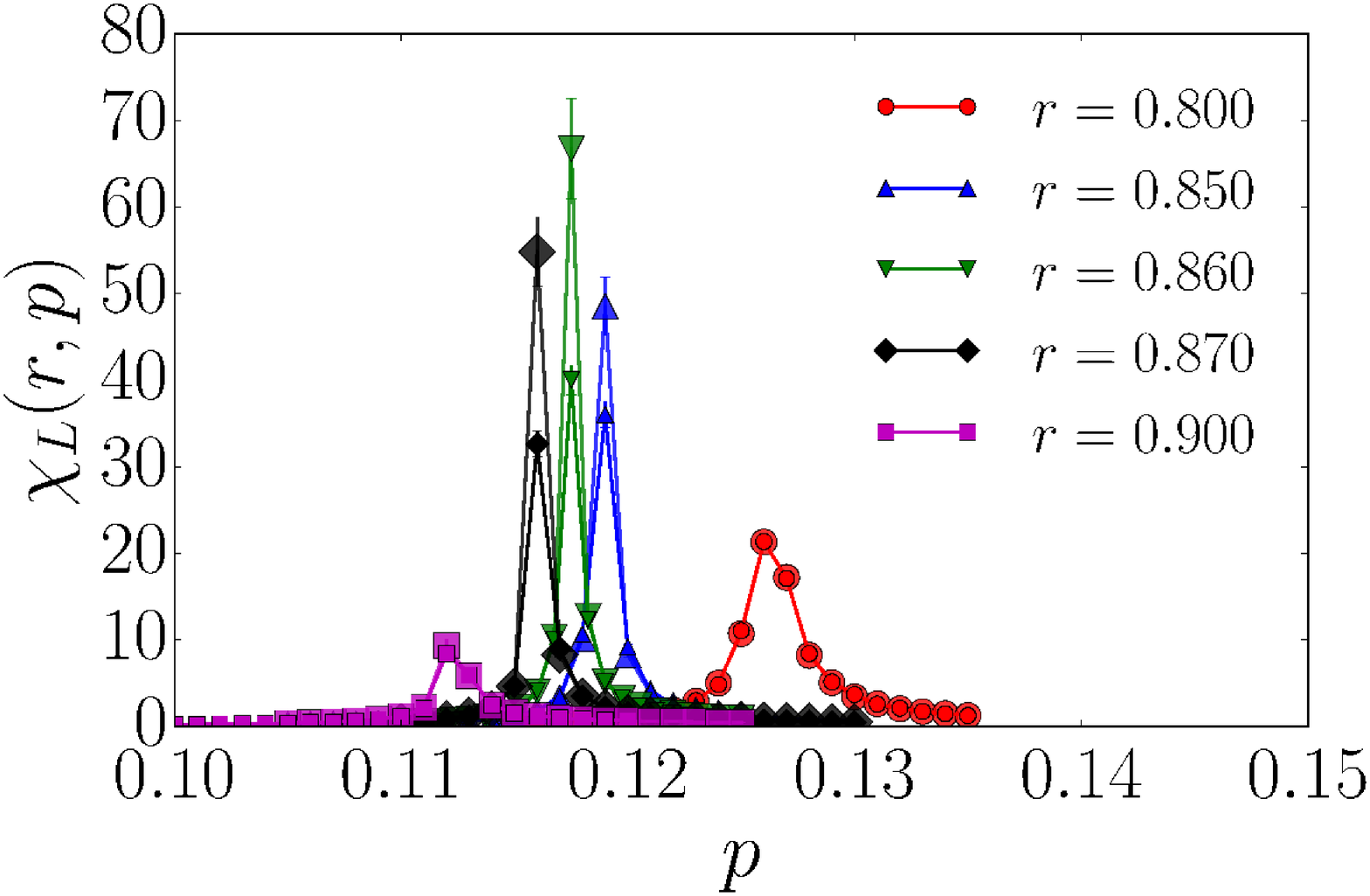}
\end{minipage}
  \caption{\textbf{Coexisting states and fluctuations on a square lattice for $m=1$.} (left) Measuring the region of coexisting states by running through a hysteresis-like loop on a square lattice with $N=2048\times 2048$ nodes for fixed $r$ and varying $p$. (right) Fluctuations of the spontaneous recovery model for different fixed values of $r$ and varying $p$. At around $r=0.86(1)$ and $p=0.117(3)$ we find the largest fluctuations corresponding to the bifurcation point. Simulations were performed for $N=128\times 128$ (smaller symbols) and $N=256\times 256 $ (larger symbols) nodes (50 samples).} 
 \label{fig:fluctuations_cp}
\end{figure}
In the square lattice we search for the bifurcation point by first analyzing the hysteresis behavior of the dynamics as shown in Fig. \ref{fig:fluctuations_cp} (left). The region where the area defining the multiple states in the hysteresis curve becomes negligible characterizes the vicinity of the cusp point. We then search for the critical point by measuring the fluctuations in that region:
\begin{equation}
\chi_L(r,p) = L^2 \left[\langle a_{st}^2\rangle-\langle a_{st}\rangle^2\right].
\end{equation}
The fluctuations in the vicinity of the bifurcation point are shown in Fig. \ref{fig:fluctuations_cp}. We conclude that the cusp point where both spinodals meet is located around $(r_c,p_c)=(0.86(1),0.117(3))$. 
%The critical behavior at this point is illustrated in Fig. \ref{fig:exponents_m1}. Although, the exponents are close to the ones of the two-dimensional Ising model 

In summary, the arguments in Appendix \ref{sec:connection} and above for the case $m=3$ (analytical and numerical) show the similarities between our model and the (non-equilibrium) contact process belonging to the directed percolation universality class \cite{henkel08}. Thus, we do not expect the dynamics to belong to the Ising universality class as conjectured in Ref. \cite{majdandzic14}. As already mentioned in Ref. \cite{grassberger82} if this contact process dynamics would belong to the Ising universality class it would mean the extension of the universality hypothesis from models with detailed balance to models without it.

%In addition, the exponents were measured at the bifurcation point, different from the percolation critical point as described in Sec. Appendix \ref{sec:connection}.

%
%
%
%\begin{figure}
%\begin{minipage}{0.49\textwidth}
%\centering
%\includegraphics[width=\textwidth]{imgs/exponent_beta.pdf}
%\end{minipage}
%\begin{minipage}{0.49\textwidth}
%\centering
%\includegraphics[width=\textwidth]{imgs/exponent_delta.pdf}
%\end{minipage}
%  \caption{\textbf{Critical exponents of the square lattice with $m=1$.} Simulation of the spontaneous recovery model with $q=1.0$, $q'=0.1$ and $m=1$.  (left) The order parameter $\Delta a_{st} (r,p_0)$ in the vicinity of the critical point $(r_0,p_0)=(0.86(1),0.117(3))$ for different $r$. (right) The order parameter $\Delta a_{st} (r_0,p)$ at critical point for varying $p$. The simulations have been performed on a square lattice with $N=1024\times 1024$ nodes. } 
% \label{fig:exponents_m1}
%\end{figure}

%
%
%
\end{document}